\documentclass[preprint,amsmath,amssymb]{revtex4-1}
\usepackage[pdftex]{graphicx}

\begin{document}
\title{Soliton dynamics in the multiphoton plasma regime}
 
\author{Chad A. Husko$^{1}$, Sylvain Combri\'{e}$^2$, Pierre Colman$^2$, Jiangjun Zheng$^{1}$, Alfredo De Rossi$^2$, Chee Wei Wong$^{1}$}
\affiliation{$^1$Optical Nanostructures Laboratory, Columbia University New York, NY 10027 USA \\ $^2$ Thales Research and Technology, Route D\'epartementale 128, 91767 Palaiseau, France} 

\renewcommand{\abstractname}{} 
\begin{abstract}
\vspace{0.5cm}
\textbf{Solitary waves have consistently captured the imagination of scientists, ranging from fundamental breakthroughs in spectroscopy and metrology enabled by supercontinuum light, to gap solitons for dispersionless slow-light, and discrete spatial solitons in lattices, amongst others. Recent progress in strong-field atomic physics include impressive demonstrations of attosecond pulses and high-harmonic generation via photoionization of free-electrons in gases at extreme intensities of 10$^{14}$ W/cm$^2$. Here we report the first phase-resolved observations of femtosecond optical solitons in a semiconductor microchip, with multiphoton ionization at picojoule energies and 10$^{10}$ W/cm$^2$ intensities. The dramatic nonlinearity leads to picojoule observations of free-electron-induced blue-shift at 10$^{16}$ cm$^{-3}$ carrier densities and self-chirped femtosecond soliton acceleration. Furthermore, we evidence the time-gated dynamics of soliton splitting on-chip, and the suppression of soliton recurrence due to fast free-electron dynamics. These observations in the highly dispersive slow-light media reveal a rich set of physics governing ultralow-power nonlinear photon-plasma dynamics.}
\end{abstract}

\maketitle
 
\section{Introduction}
Recent advances in nonlinear optics have seen rapid developments spanning from single cycle plasma interactions \cite{goulielmakis2008single}, chip-scale parametric frequency conversion \cite{combReview,osgood_midIR,corcoranTHG}, to slow-light enhanced nonlinearities \cite{eggletonNaturePhysics2006,baba_nature2008}, discrete spatial solitons\cite{fleischer2003}, and temporal cloaking \cite{fridman2012} in the past few years. Solitons are a special class of nonlinear waves arising from the interplay of dispersion and nonlinear effects \cite{dauxoisBook,ilan2010band,kivsharBook}. Soliton-based phenomena give rise to optical rogue waves \cite{jalaliNature07}, pulse compression \cite{gaetaScience03}, Raman-dispersive wave interaction \cite{skryabin2003soliton}, self-similarity \cite{dudleyNP2007}, and supercontinuum optical sources, enabling key applications in spectroscopy and metrology \cite{dudleyRMP}.
 
\indent
In parallel to these developments, strong-field atomic physicists have adopted concepts from the plasma community, leading to powerful physical insight and subsequent demonstrations of attosecond pulses and high-harmonic generation via photoionization of free-electrons in gases \cite{goulielmakis2008single,macklin1993high,corkum1993plasma}. Many of these experiments focus on the tunneling regime of atomic gases \cite{macklin1993high, goulielmakis2008single,DiMauroNP2010} with guided wave tunnel ionization of noble species only recently demonstrated \cite{holzer2011ionization,fedotov2007ionization}. Exploration of the complementary process of multiphoton plasma generation often involves ultraviolet and extreme ultraviolet sources with complex detection schemes \cite{goulielmakis2008single,mcpherson1987studies}. These experiments, moreover, typically occur at 10$^{14}$ W/cm$^2$ intensity levels for sub-100 fs pulses. 
 
\indent
The semiconductor chip-scale platform alternatively presents large nonlinearities along with strong field localization to enable record low power observations and scalable optoelectronic integration. Recent efforts on ultrafast pulses in microchip devices include the first temporal measurements of solitons on-chip~\cite{colman2010,tanNComms2010}, inferred indirectly through intrinsically-symmetric intensity correlations \cite{karle2011direct} or estimated from spectral measurements~\cite{ding2010time}. 

Here we report the first phase-resolved spectroscopy of ultrafast optical solitons in slow-light photonic crystals. The optically-gated spectrograms evidence the first observations of: (1) chip-scale dynamical soliton pulse splitting with temporally-flat phase, (2) self-induced pulse acceleration due to multiphoton carrier plasma and non-adiabatic chirp, and (3) suppression of soliton recurrence due to fast free-electron dynamics in our GaInP $\chi^{(3)}$ media. The strong light confinement and light-matter interaction enable the observations at $\sim$ 10 pJ and picosecond pulses, yielding $\sim$ 10$^{10}$ W/cm$^2$ intensities, in a 1.5-mm photonic lattice. The novel coupled soliton-plasma dynamics in the semiconductor are rigorously examined in a modified nonlinear Schr\"odinger framework including auxiliary carrier evolution, providing strong agreement based on experimentally measured parameters without any fitting. The observations are described by uniting concepts of ultrafast nonlinear solitons, high-field atomic physics, and semiconductor physics. Twelve sets of dispersive propagation conditions are characterized and a scaling law derived for soliton compression on-chip, incorporating Kerr, three-photon absorption and free carrier nonlinearities, and slow-light dispersive characteristics. A minimum pulse duration of 440 fs is achieved in our higher-order soliton compression with a 20.1 pJ, 2.3 ps input pulse, exhibiting a precise phase balance between slow-light-enhanced Kerr self-phase modulation and strong group velocity dispersion in our microchip. Beyond these measurements, our approach provides an exploration into a new regime of light-plasma interaction.
 
\section{Soliton dynamics in the multiphoton plasma regime}
\subsection{Sample group velocity dispersion and frequency-resolved optical gating}

\begin{figure}
\centering
\includegraphics[height=4.6cm]{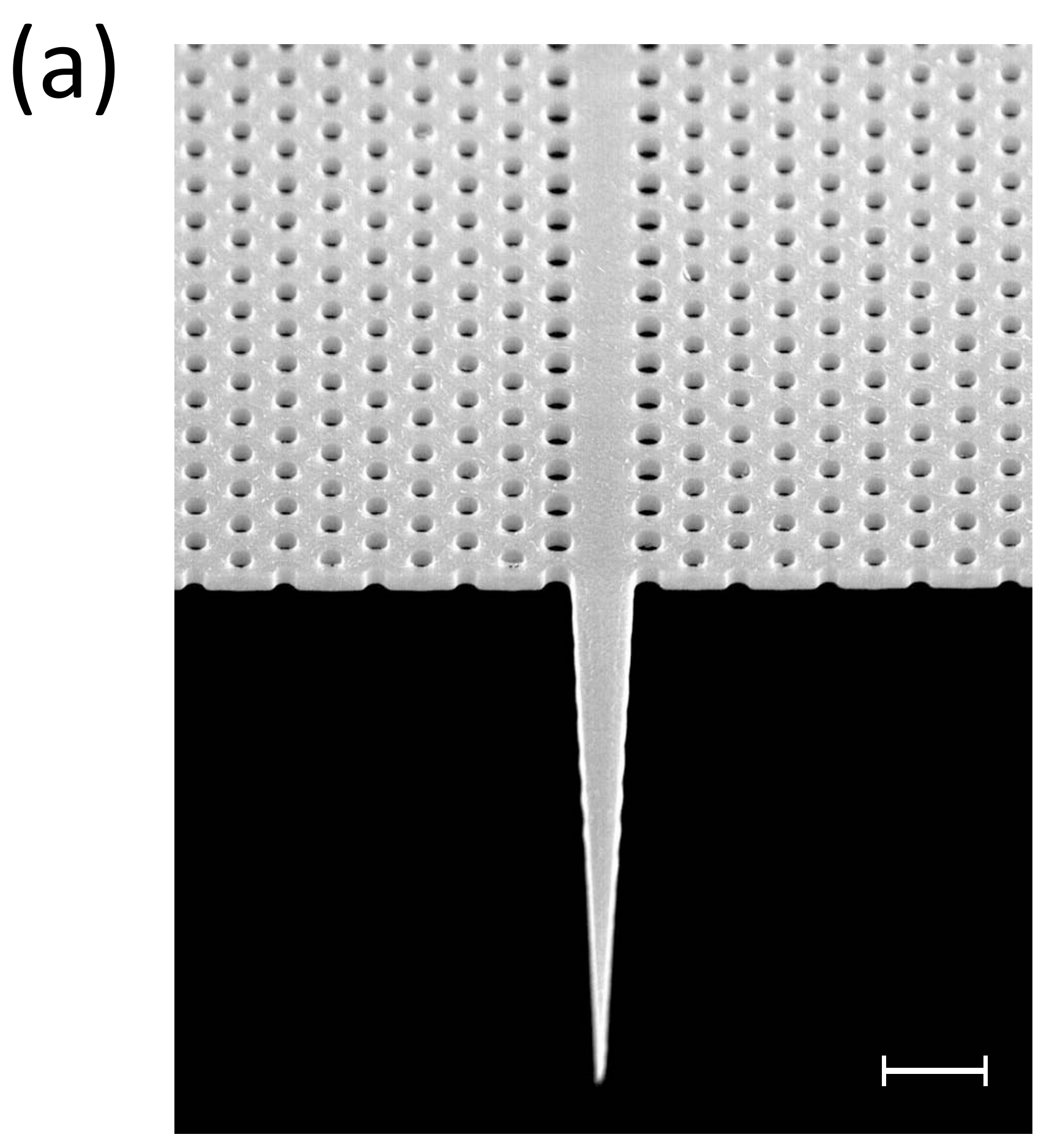}
\includegraphics[width=6cm]{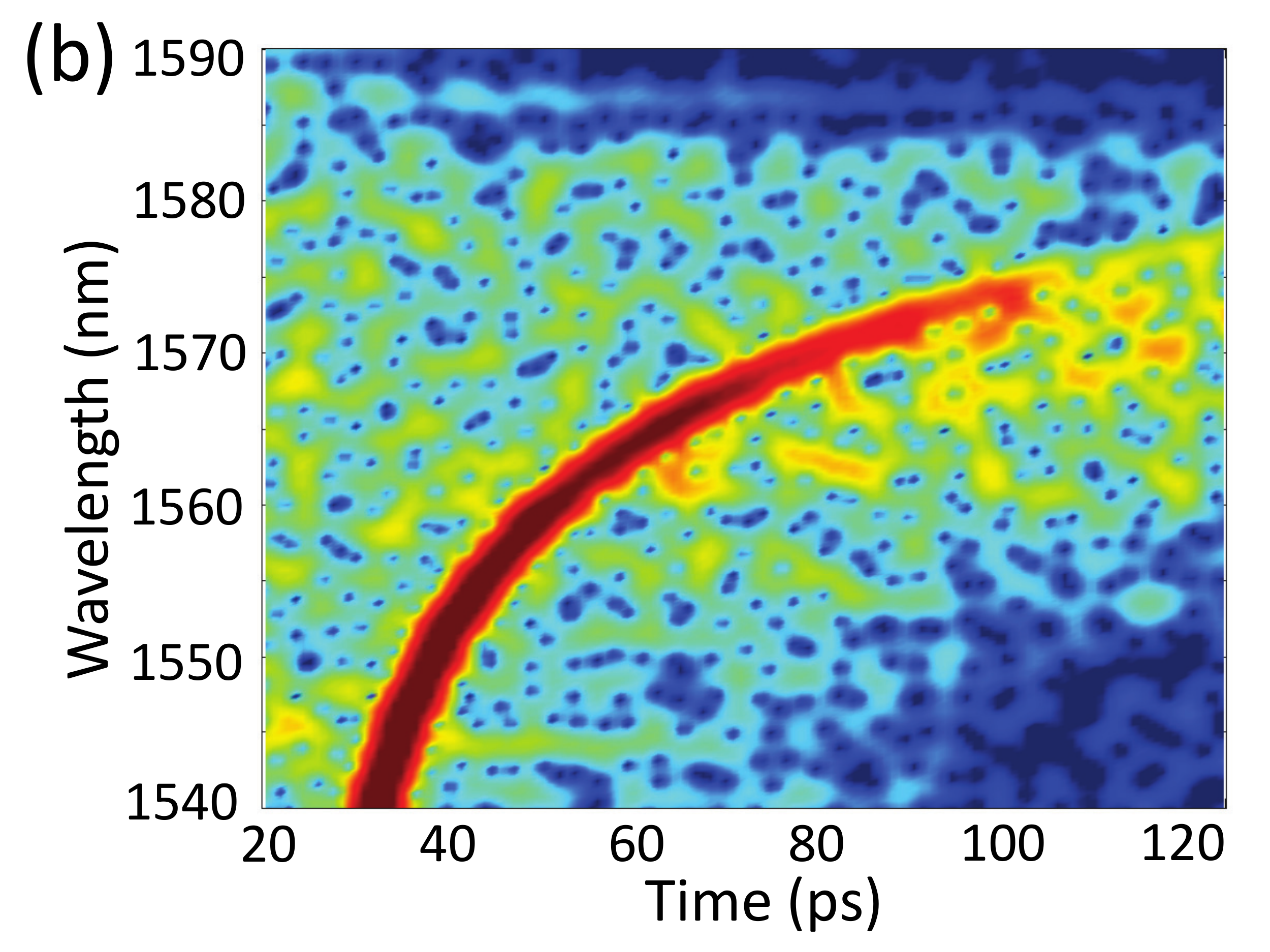}
\includegraphics[width=6cm]{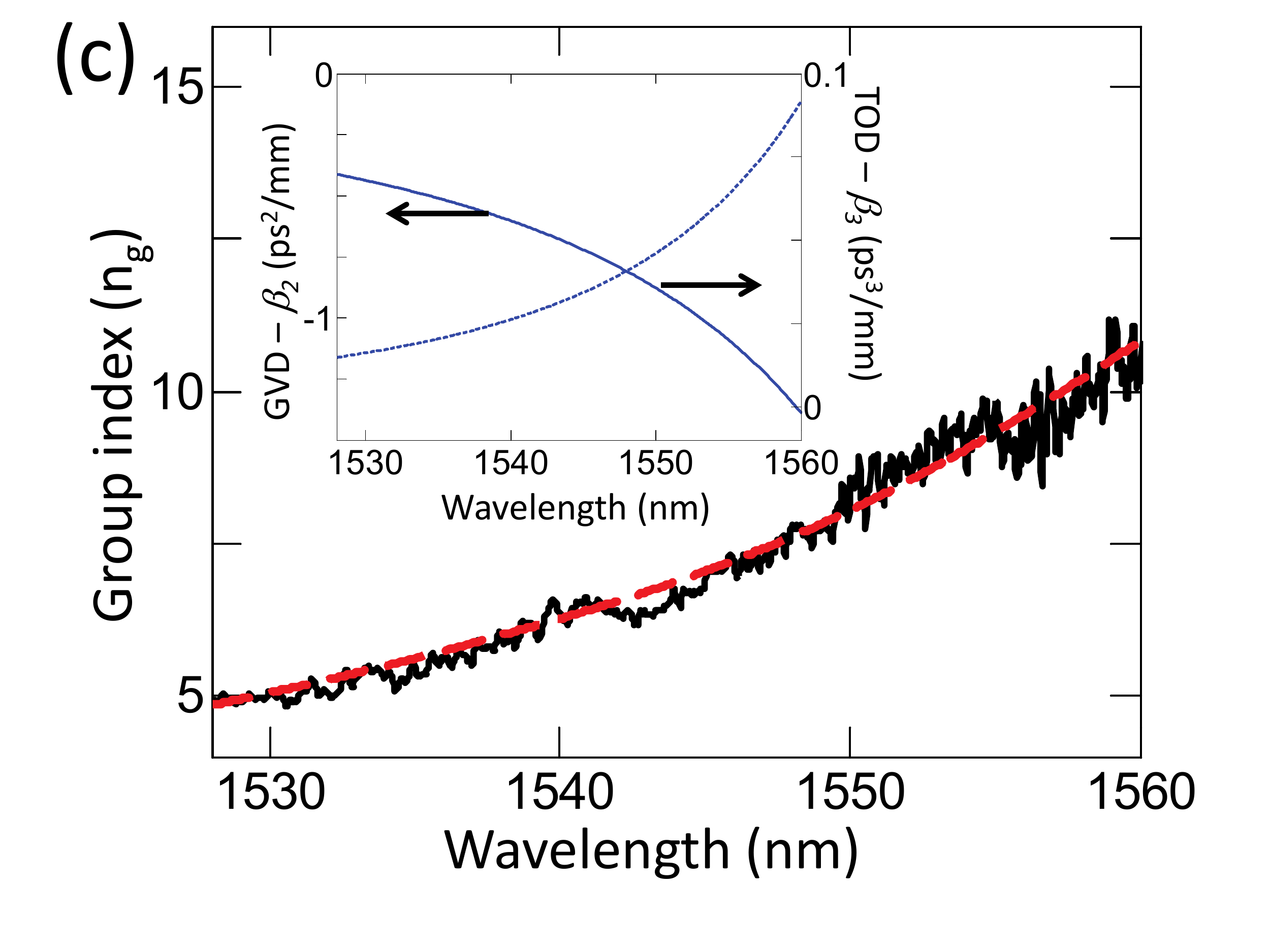}
\caption{Ultrafast soliton dynamics in photonic crystal microchip. (a) Scanning electron micrograph of GaInP membrane with designed mode adapters (Scale: 1 $\mu$m) \cite{tranAPL09}. (b) Waveguide dispersion properties measured via the phase-shift method. (c) Measured group index (solid black) with the phase-shift technique \cite{combrie2006} and calculations used to compute the dispersion and modal area (dashed red) \cite{mpbCode}. Inset: Group velocity dispersion (left axis) and third order dispersion (right axis) derived from first-principle band structure computational comparison to dispersion measurements.
}
\label{fig:figure_linParams}
\end{figure}

\indent
Figure \ref{fig:figure_linParams}(a) illustrates the GaInP photonic crystal membranes examined, with a hexagonal lattice constant \textit{a} of 475 nm, 0.18\textit{a} hole radius and a 195 nm thickness, and a line-defect dielectric that forms the photonic crystal waveguide. The dispersion is engineered by tuning the innermost hole radii to 0.21\textit{a}. The 1.5-mm photonic crystal waveguide includes integrated mode-adapters \cite{tranAPL09} to reduce the total input-output chip insertion losses to $\sim$ 13 dB and suppress facet Fabry-Perot oscillations (see Supplementary Information). Figure \ref{fig:figure_linParams}(b) indicates the waveguide dispersion properties measured via the phase-shift method. Figure \ref{fig:figure_linParams}(c) shows the group index, $n_g$, increasing from 5 to 12 in the range of interest. The dashed line indicates band structure calculations employed to compute the dispersion and modal area \cite{mpbCode}. Figure \ref{fig:figure_linParams}(c)(inset) shows the group-velocity dispersion (GVD, $\beta_2$) and third-order dispersion ($\beta_3$) coefficients of the device. The group velocity dispersion is anomalous and on the order of $\sim$ ps$^2$/mm across the range of interest. We emphasize third-order dispersion plays a negligible role here, and is included for completeness. The GaInP material selection has negligible red-shift Raman effects in contrast to solitons in amorphous materials such as glass. Moreover, in contrast to nonlinear waves in silicon \cite{tanNComms2010,ding2010time,zhang2007optical,monat2009} where two-photon absorption greatly restricts the full range of dynamics,  GaInP has a large 1.9-eV band gap to completely suppress any two-photon absorption (of 1550-nm photons) and has negligible residual effects from band tail absorption \cite{husko2009_OptExp}. The three-photon GaInP material employed \cite{combrie2009} enables the fine balance between the soliton propagation and plasma regimes.

\indent 
Photo-induced plasmas are characterized by the Keldysh parameter, $\kappa = \sqrt{\frac{I_p}{2U_p}}$, where $I_p$ is the ionization potential and $U_p$ the ponderomotive energy \cite{keldysh1965ionization}. $\kappa>1$ defines the multiphoton regime, whereas $\kappa<1$ corresponds to a tunneling dominated process. In the experiments presented here $\kappa \approx$ 5 - 6, well into the multiphoton plasma regime. This is largely due to the four orders of magnitude reduced intensities (10$^{10}$ W/cm$^2$) required to ionize the semiconductor media compared to gases (10$^{14}$  W/cm$^2$) \cite{macklin1993high, goulielmakis2008single,DiMauroNP2010,holzer2011ionization,fedotov2007ionization}.
 
\indent 
For ultrashort pulse characterization we constructed a 25 fJ phase-sensitive second-harmonic-generation (SHG) FROG apparatus (see Methods). Frequency-resolved optical gating (FROG) \cite{trebinoBook, thesisXing} or spectral-phase interferometry \cite{pasquaziNP2011} enables the complete pulse intensity and phase retrieval in both spectral and temporal domains, covering supercontinuum \cite{dudleyRMP} and attosecond \cite{DiMauroNP2010, mairesse2005atto} pulse regimes. In order to guarantee fidelity of the pulses collected off-chip, our experiments with the cryogenic detectors exclude erbium-doped fiber amplifiers and are externally intensity-attenuated to avoid any modification of the pulse properties.
\subsection{Soliton dynamics in the presence of plasma: acceleration and temporal splitting}
 
\begin{figure}[*h]%
\centering
\includegraphics[height=13cm]{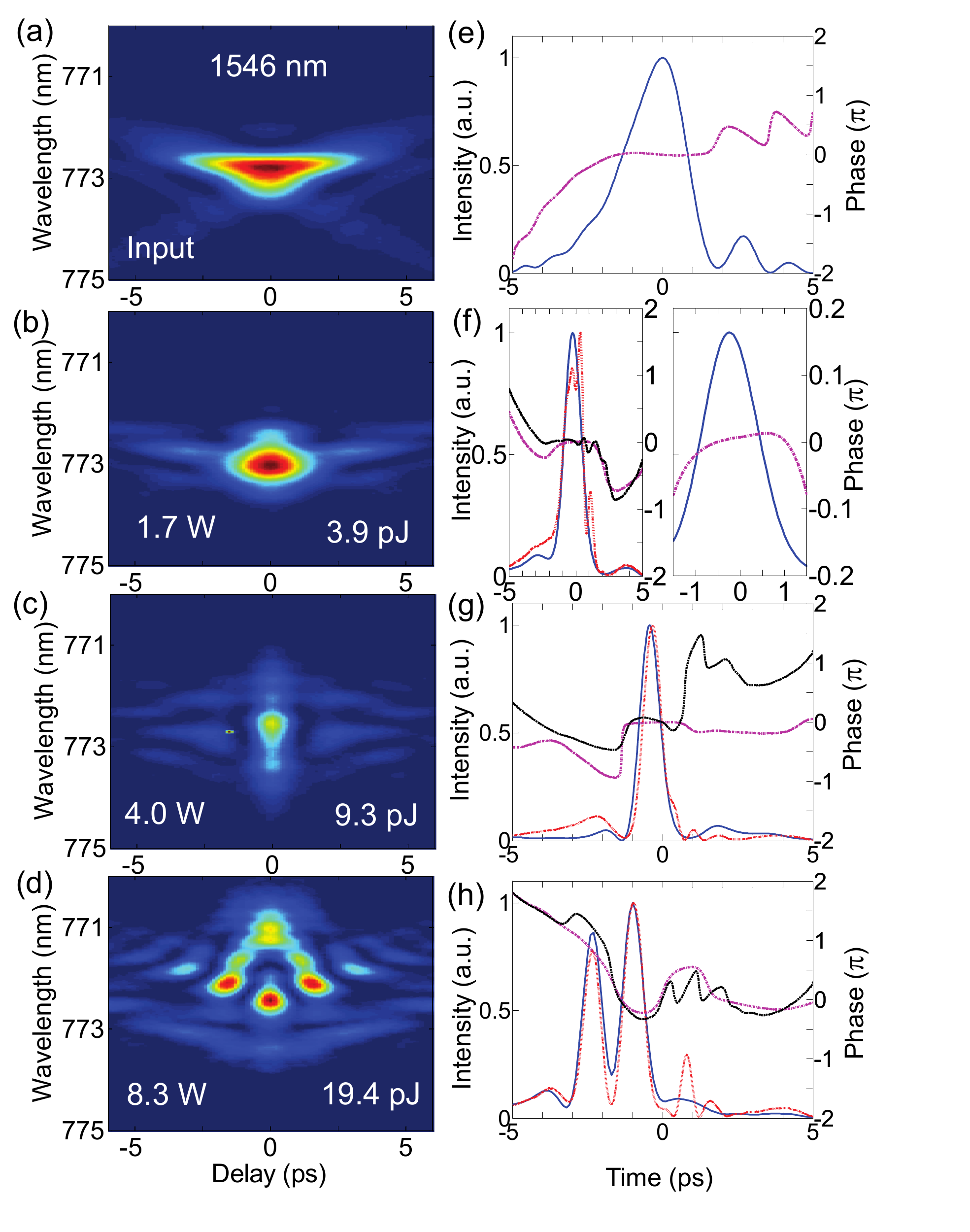}
\includegraphics[height=13cm]{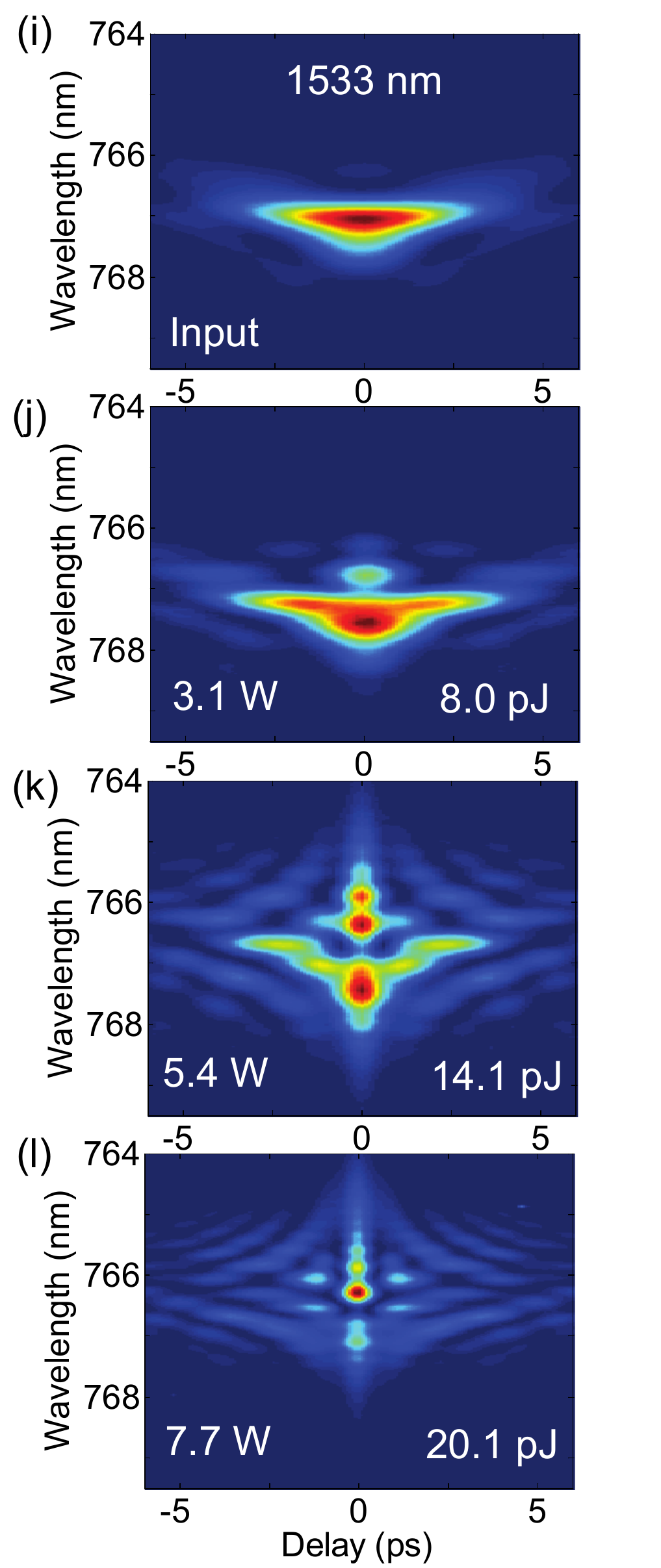}
\caption{Frequency-resolved optical gating of solitons in highly-dispersive photonic crystal waveguides. (a)-(d): FROG spectrograms with coupled pulse energies from 3.9 pJ to 19.4 pJ, with input pulses centered at 1546 nm. (e)-(h): FROG retrieved time domain intensity (solid blue) and phase (dashed magenta), with gating error less than 0.005 on all runs. Superimposed nonlinear Schr\"odinger equation modeling: intensity (dashed red), and phase (dash-dot black), demonstrates strong agreement with experiments. The right side of Panel (f) is a zoom of the FROG data demonstrating the flat soliton phase within  0.1 radians or less. Panels (c) and (g): The pulses first compress to a minimum duration of 770 fs at 4 W (9.3 pJ), before splitting into two peaks at 8.3 W (19.4 pJ) [panels (d) and (h)], exhibiting higher-order soliton dynamics.(i)-(l): FROG spectrograms with input pulses centered at 1533.5 nm.}
\label{fig:frogTraces}
\end{figure}

 \indent
First we characterized the pulse evolution in the waveguide with FROG as a function of input pulse energy for a broad array of dispersion and nonlinear properties. We highlight three cases demonstrating the unique aspects of nonlinear pulse evolution. We first focus on the 1546 nm case ($n_g$ = 7.2, $\beta_2$= -0.75 ps$^2$/mm), near the band edge, which exhibits the greatest diversity of nonlinear pulse dynamics. Figs. \ref{fig:frogTraces} (a)-(d) show experimental FROG traces of the input pulse and at three different pulse energies. Figs. \ref{fig:frogTraces} (e)-(h) immediately to the right are the retrieved temporal intensity (solid blue) and phase (dashed magenta) of the FROG traces, with retrieved optical gating errors less than 0.005 in all cases demonstrated (see Methods). Fig.~\ref{fig:frogTraces}(i)-(l) are the corresponding 2D spectrograms for pulse centered at 1533.5 nm. The spectral properties exhibit higher-frequency components generated by the free-electrons. Fig. \ref{fig:frogTraces}(f) shows the maximum temporal compression to 770 fs at 1546 nm (compression factor $\chi_c = T_{in}/T_{out}=2.8)$. The right panel shows a magnified view of the output pulse phase which illustrates the phase is flat and uniform within 0.1 radians or less, confirming for the first time the presence of the chip-scale optical soliton with its phase in the highly-dispersive nonlinear media.
 
\indent
Next we examine higher-order soliton evolution near the photonic crystal band edge. Soliton propagation is determined by two length scales \cite{agrawalNLoptics}, the nonlinear length $L_{NL}= 1/(\gamma_{eff} P_o$) [with the effective nonlinear parameter $\gamma_{eff}$ defined by $\frac{n_2k_0}{A_{3eff}}(\frac{n_g}{n_0})^2$\cite{husko2009_OptExp} and $P_o$ as the pulse peak power] and the dispersion length $L_D= T_o^2/\beta_2$ [where $T_o=T/\Gamma$, $T_o$ as the pulse width (FWHM), here $T$=2.3 ps, and $\Gamma = 2\cosh^{-1}(\sqrt{2}) = 1.76$ for hyperbolic secant pulses]. The soliton number $N=\sqrt{L_d/L_{NL}}$ defines the conditions for soliton propagation. Let us first consider the canonical case of GVD and SPM only, i.e. neglecting higher-order effects. When $N$=1 in this simple case, the pulse propagates without dispersing as a fundamental soliton due to a precise balance of GVD and SPM. When $N>$1 in the simple case, the higher-order solitary pulse evolves recurrently by first compressing, then splitting temporally before regaining its initial shape after a soliton period $z_o = \frac{\pi}{2} L_d$. In contrast to these simple dynamics, in our semiconductor media the soliton propagation dynamics are governed by a complex nonlinear regime involving an intrapulse non-adiabatic free-carrier plasma (with absorptive and dispersive terms) generated from three-photon absorption, giving rise to the composite temporal and spectral features in the 2D spectrograms of Figure~\ref{fig:frogTraces}.
 
\indent
To discern the roles of each of these effects, we model the pulse propagation with a nonlinear Schr\"odinger equation (NLSE) \cite{Bhat_PRE2001} that captures the underlying perturbed Bloch lattice with an envelope function, including dispersive slow light, free-carrier dynamics (density $N_c$), three-photon absorption, and higher-order effects. The full model employed here (detailed in the Methods) contrasts with the simple integrable NLSE which neglects losses, gain, and any higher-order dispersion. Importantly, free-electrons exhibit distinct ionization and loss dynamics in the two plasma (multiphoton versus tunneling) regimes. Here in the multiphoton regime the free-electron absorption is proportional to $N_c$, whereas in the tunneling regime (not present here), loss is proportional to the ionization rate $dN_c/dt$ \cite{saleh2011theory,fedotov2007ionization}. These dynamics are included in the model. The high-sensitivity FROG captures the exact pulse shape and phase of the input pulses, which subsequently serves as the initial launch pulse conditions into the NLSE. The resulting NLSE predicted intensity (dashed red) and solitary phase (dash-dot black) are presented in Figs. \ref{fig:frogTraces}(f)-(h). Since FROG only gives the relative time, we temporally offset the FROG traces to overlap the NLSE for direct comparison. With all parameters precisely determined from experimental measurements, e.g. with no free parameters, we observe a strong agreement between the femtojoule-resolution measurements and the NLSE model across the diverse array of pulse energies and center frequencies.
 
\indent
Examining further the soliton temporal dynamics, we illustrate both the FROG and NLSE at 1546 nm and 1533.5 nm for varying pulse energies in Fig. \ref{fig:pulseTimeShift}(a). As we increase the pulse energy, the center of the pulse, defined by the first-order moment, forward shifts from 0 ps to -1.4 ps, indicating for the first time that the 19.4 pJ (8.3 W) pulse accelerates as it travels along the slow-light photonic crystal. The black dashed line indicates simulations with suppressed free-carrier effects ($N_c$ =0) at 19.4 pJ (8.3 W). The pulse center shifts noticeably less in this case, with a difference of 0.68 ps, confirming the origin of the soliton acceleration and its accompanying blue-shift is free-carrier plasma. Fig. \ref{fig:pulseTimeShift}(b) shows the modeled pulse intensity and generated carrier population along the waveguide, with these frequencies near the band edge. Fig. \ref{fig:pulseTimeShift}(b) indeed indicates the self-induced blue-shift -- or a self-induced frequency-chirp -- is strongest near the input of the photonic crystal and at the location of largest pulse compression, correlating with the locations of highest powers and therefore generated free-electrons. These effects clearly arise from the non-adiabatic generation of a carrier plasma within the soliton itself. We additionally confirmed that third-order dispersion is negligible (not pictured here) in this regime by comparing it on and off in the model, indicating the acceleration is due completely to the generated plasma. The suppressed third-order-dispersion model further indicates the residual temporal shift is due to a small initial chirp in the pulses. This is in stark contrast with above-band-gap carrier-injection derived from adiabatic processes that shift the bands themselves (see Supplementary Information) \cite{leonard2002ultrafast,kampfrath2010ultrafast}.

\indent
Next we tune the soliton frequencies further away from the band edge, with an example 1533.5 nm case ($n_g$ = 5.4, $\beta_2$=~-0.49 ps$^2$/mm) shown in Fig. \ref{fig:pulseTimeShift}(c). Though the input pulses are nearly identical, the pulse evolution is distinct due to a reduced dispersion and faster group velocity, and therefore weaker nonlinear and free-carrier plasma effects, compared to the 1546 nm case. Full phase retrieval of the 2D spectrograms at 1533.5 nm, similar to Fig. \ref{fig:frogTraces}, are detailed in the Supplementary Information. Examining the temporal intensity, Fig. \ref{fig:pulseTimeShift}(c) indicates a minimum duration of 440 fs ($\chi_c=5.2)$, 330 fs shorter than the 1546 nm ($N$ =2.4) case due to the larger injected soliton number ($N$=3.5) for optimal compression at this dispersion and sample length. The 1533.5 nm pulse also experiences less acceleration due to smaller self-induced frequency-chirp and free-carrier effects, as illustrated in Fig. \ref{fig:pulseTimeShift}(d), confirming the robustness of the soliton acceleration mechanism under different nonlinear, dispersion, and input pulse conditions.

\subsection{Periodic soliton recurrence and suppression in the presence of free-electron plasma}

\indent 
Dynamical solitons, in the balance of Kerr nonlinearity with anomalous dispersion, exhibit periodic recurrence -- the soliton breakup, collision and re-merging \cite{kiblerNature2012,wuPRL2007} --  in a Fermi-Pasta-Ulam lattice. Here we examine soliton dynamics at 1555 nm ($n_g$ = 9.3, $\beta_2$ = -1.1 ps$^2$/mm) for the higher-order solitons in the presence of free-carriers and nonlinear absorption. Figure \ref{fig:carrierEffects}(a) shows the experimentally captured FROG trace at 14.8 pJ (6.3 W) alongside NLSE modeling of the pulse propagation, including auxiliary free-electron non-instantaneous dynamics. The higher-order ultrafast soliton evolves by first compressing to a minimum duration, then splitting temporally, with the accelerated pulse induced by the multiphoton plasma. Fig. \ref{fig:carrierEffects}(b) shows the measured temporal trace in comparison with the NLSE simulations at the waveguide output, with remarkable agreement between experiment and theory.
 
\indent
Based on the high-fidelity of our model, we next consider numerically the case of a sample with twice the length $L'=2L$, such that $L'>z_0$. Given the input soliton number of $N$=3.2, we expect the pulse to have nearly reformed since the simulated sample length $L' = 3$ mm is greater than the soliton period $z_0 = 2.7$ mm \cite{agrawalNLoptics}. Fig. \ref{fig:carrierEffects}(c), however, clearly demonstrates irreversible blue-shifting of the pulse energy, thereby breaking the symmetry of the optical pulse periodic evolution such that soliton recurrence is not possible in the presence of a plasma. The dominant loss mechanism at large peak powers is three-photon absorption, with a much smaller contribution from free-carrier absorption. Further to this point, in Fig. \ref{fig:carrierEffects}(d) we illustrate the pulses with suppressed free-carrier effects ($N_c$=0) while retaining three-photon absorption and an ideal hyperbolic secant input. These figures exhibit recovery of the pulse symmetry without temporal shifts, illustrating the sizable contribution of the multiphoton plasma to the nonlinear dynamics. Though the temporal shape is symmetric, the higher-order periodic evolution is suppressed in this case due to three-photon absorption lowering soliton number $N<1$. Indeed here three-photon absorption is the ultimate limit to loss in the multiphoton regime, in contrast to carrier generation dominating in the tunneling regime. The strong contrast of panel (d) with the other panels demonstrates the suppression of periodic recurrence in the presence of free-carrier dynamics. We note that materials with large two-photon absorption such as silicon cannot exhibit these dynamics at similar wavelengths and across the full range of pulse energies. 

\subsection{Temporal compression in multiphoton plasma materials} 
\indent
In soliton pulse compression schemes, it is important to consider the point of optimal temporal narrowing, $z_{opt}$. Fig. \ref{fig:minimumDuration}(a) highlights the ratio of optimal length $z_{opt}$ to soliton period $z_0$ versus soliton number $N$ in GaInP computed via NLSE. The results include the slow-light modified Kerr, three-photon and free-carriers, and can be cast by the following fitted relation: 
\begin{equation}
\label{zOptAnalytic}
\frac{z_{opt}}{z_0} = \frac{0.7}{N} - \frac{0.9}{N^2} + \frac{4}{N^3}.
\end{equation}
In the experimental case of 1546 nm maximum compression was achieved at $N=2.4$, corresponding to $z_{opt}/z_0 \approx$ 0.37. This yields an estimate of $z_{opt}$ of 1.37 mm, in solid agreement with the effective sample length of $L_{eff}$ of 1.35 mm at this wavelength. To discern the role of the nonlinear effects $n_{2eff}$ and $\alpha_{3eff}$ in the compression dynamics, we investigate a 50\% larger $\alpha_{3eff}$ (same base $n_{2eff}$) and $n_{2eff}$ (same base $\alpha_{3eff}$), indicated by the lines above and below the experimental case, respectively. As shown, a larger effective $n_{2eff}$ ($\alpha_{3eff}$) causes the $z_{opt}/z_0$ vs. $N$ curve to move downwards (upwards), e.g. decreasing (increasing) the length scale of compression, and indicating that desired compression effects can be achieved at lower (higher) intensities. Thus different effective nonlinearities will have different scalings due to enhanced or suppressed compression dynamics. For soliton compression in semiconductor media it is clearly important to consider the balance between Kerr and nonlinear absorption.
 
\indent
We next carried out measurements to determine the minimum pulse duration for twelve different wavelengths, mapping the dispersion conditions across a broad range of slow group velocity regions. Fig. \ref{fig:minimumDuration}(b) summarizes these results at the achieved compression factor $\chi_c$ versus the measured soliton number $N$ at the minimum temporal duration. At slower group velocities (longer wavelengths), the ultrafast compression scales monotonically with $N$, along with the minimum pulse duration approaching 440 fs from a 2.3 ps pulse input at 1533.5 nm (input pulse parameters detailed in the Supplementary Information). Larger $N$ values corresponds to greater compression factors, as expected, with all of the wavelengths examined experiencing a compression of at least $\chi_c>2$. The principles of dispersion engineering could allow for uniform $N$ and $\chi_c$ to create broadband soliton compression \cite{baba_nature2008,corcoranTHG}.

%
================================================================
\section{Conclusion} 
We have demonstrated soliton dynamics in the multiphoton plasma regime in highly-nonlinear, highly-dispersive, photonic crystal lattices. We observed phase-balanced optical solitons, dynamical pulse splitting, solitary pulse acceleration due to self-induced frequency-chirp, in addition to the suppression of soliton recurrence due to fast carrier dynamics via frequency-resolved optical gating spectroscopy. Higher-order soliton compression down to 440 fs from 2.3 ps was observed at 20.1 pJ in 1.5-mm device lengths. We characterized soliton compression at twelve sets of dispersion values and derived a scaling for compression and soliton number in semiconductors. The demonstrated ultra-low energies (10s of pJ) and intensities ($\sim$ 10$^{10}$ W/cm$^2$) are six and four orders of magnitude smaller, respectively, than required in recent amorphous materials for significant plasma photoionization and densities~\cite{holzer2011ionization,fedotov2007ionization} and even smaller than that of attosecond extreme ultraviolet radiation  in gases~\cite{goulielmakis2008single}. These observations of strong light-matter interaction at $\sim$ pJ energies in nanophotonic architectures advance our understanding of nonlinear wave propagation and open key new research pathways towards fundamental studies of multiphoton light-plasma interactions.
\\ 

\noindent \textbf{Acknowledgements}

The authors thank James F. McMillan, Jie Gao, Matthew Marko, and Xiujian Li for useful discussions, and Keren Bergman for the autocorrelator. C.A.H. performed the measurements and numerical simulations. J.Z. assisted in the building of the FROG setup. P.C. and S.C. prepared the samples and nanofabrication, and P.C. assisted in the modeling. A.D.R. and C.W.W. supervised the project. C.A.H., A.D.R., and C.W.W. wrote the manuscript. All authors confirm the advances described in the paper. The work is partially funded by the National Science Foundation, under ECCS-1102257, DGE-1069240, and ECCS-0747787, and the 7th Framework Program of the European Commission program COPERNICUS (www.copernicusproject.eu). The authors declare no competing financial interests. Correspondence and requests for materials should be addressed to C.A.H. and C.W.W.
 
\section*{Methods}
\subsection*{Experimental pulse characterization}
\indent
In the soliton measurements, we employed a mode-locked fiber laser (PolarOnyx) delivering nearly transform-limited 2.3 ps pulses at a 39 MHz repetition rate. The source is tunable from 1533.5 to 1568 nm. We characterized the input pulses with the FROG, experimentally verifying that the time-bandwidth product approaches the Fourier-limit of hyperbolic secant pulses ($\Delta \lambda \Delta \nu$= 0.315) within 5\%. The power input into the photonic crystal waveguide is attenuated with a polarizer and half-wave plate, thereby preventing misalignment and undesirable modification of the pulse shape. Importantly, the pulses collected from the end facet of the photonic crystal waveguide were input directly into the FROG to guarantee accurate measurement of the pulse, e.g. no amplification stage. 
 
\indent
The second-harmonic (SHG) FROG apparatus consisted of a lab-built interferometer with a thin BBO crystal (1 mm) and a high-sensitivity grating spectrometer (Horiba) with a cryogenically-cooled deep-depletion 1024 $\times$ 256 Si CCD array. The spectral resolution $\Delta\lambda$ was 20 pm while the delay time step $\Delta T$ was varied between 100 to 200 fs, depending on the pulse duration. The FROG can detect pulses as little as 25 fJ pulse energies (1 $\mu$W off-chip average power). The results were computed on a 256 $\times$ 256 grid and with retrieved FROG gate errors $G$ below 0.005 in all cases reported here. The FROG algorithm retrieves the pulse temporal and spectral properties, including a direct determination of the phase without any approximations. The FROG output data were compared with an optical spectrum analyzer (OSA), to ensure robust pulse retrieval of the FROG algorithm. The output pulses were too weak to measure with an autocorrelator (AC) here. The low FROG retrieval errors and good match to the spectral features indicate proper retrieval.
 
\subsection*{Nonlinear Schr\"odinger equation (NLSE) model}
\indent The NLSE model is described by \cite{Bhat_PRE2001}: 
\begin{equation}
\label{NLSE}
\frac{\partial E}{\partial z} = i\gamma_{eff}|E|^2E -i\frac{\beta_2}{2} \frac{\partial^2E}{\partial t^2}+\frac{\beta_3}{6} \frac{\partial^3}{\partial t^3}-\frac{\alpha}{2}E-\frac{\alpha_{3eff}}{2}|E|^4E +(ik_o\delta-\frac{\sigma}{2})N_cE.
\end{equation}
This includes third-order dispersion $\beta_3$, linear propagation loss $\alpha$, effective slow-light three-photon nonlinear absorption $\alpha_{3eff}$ \cite{husko2009_OptExp}, effective nonlinear parameter $\gamma_{eff}$, and generated carrier density $N_c$ with associated free-carrier dispersion $\delta$ and absorption $\sigma$. The auxiliary carrier equation induces a non-instantaneous response through the carrier lifetime $\tau_c: \frac{\partial N_c}{\partial t}=\frac{\alpha_{3eff}}{3\hbar\omega A_{3eff}}|E|^6-\frac{N_c}{\tau_c}$. The free-carrier dispersion coefficient $\delta$ includes group index scaling: $\delta = -\frac{q^2}{2\omega^2\epsilon_on_om^*}\frac{n_g}{n_o}$. Here $\sigma$ is $4\times10^{-21}(n_g/n_0)$ m$^2$ based on established data for GaAs lasers and scaled with the group index. We solve the NLSE model employing an implicit Crank-Nicolson split-step method. Parameters are obtained directly from experimental measurements or calculated as required (such as $A_{3eff}$) \cite{colman2010,mpbCode}. The bulk Kerr $n_2$ = 0.57$\times$10$^{-17}$ m$^2$/W \cite{sheikBahae1990} and $\alpha_3$ = 2.5$\times$10$^{-26}$ m$^3$-W$^{-2}$ \cite{wherrett1984} coefficients employed in the calculations are in agreement with well-known models. Third-order nonlinear effects and linear propagation loss are taken to increase with group velocity. Third-order dispersion, included in the model, contributes negligibly throughout the range of parameters examined here.
 
\subsection*{FROG characterization of launched pulses}
Before examining the soliton dynamics in the photonic crystal waveguide at various wavelengths, we first characterized the input pulse with the FROG apparatus. We observed solid agreement between the experimental and retrieved FROG traces as shown in the Supplementary Information. Comparison of autocorrelation traces between the FROG and a conventional autocorrelator (FemtoChrome) shows one-to-one matching of the launched pulses; comparison of spectral lineshapes between the FROG and an optical spectrum analyzer shows near identical matching. FROG retrieves the pulse temporal intensity and phase, information unavailable from autocorrelation or an optical spectrum analyzer alone (detailed in the Supplementary Information). The slight pulse asymmetry, for example, is obscured in the autocorrelation trace. The pulse phase is nearly flat, indicating near transform-limited performance.  Pulses at other wavelengths exhibit similar characteristics.

\clearpage
\clearpage
\clearpage
 
\begin{figure}
\centering
\includegraphics[width=14cm]{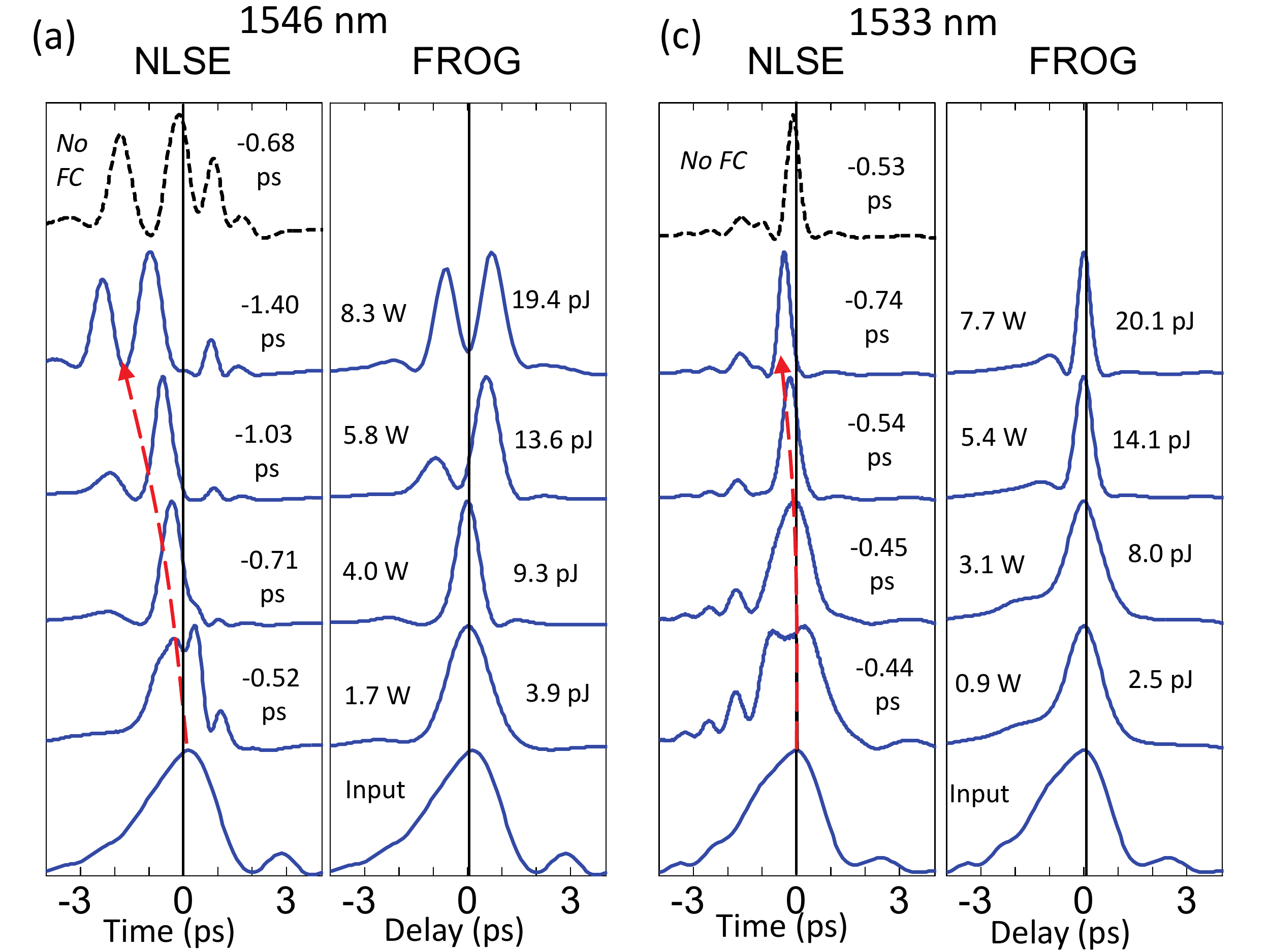}
\parbox{7.5cm}{\includegraphics[width=7.5cm]{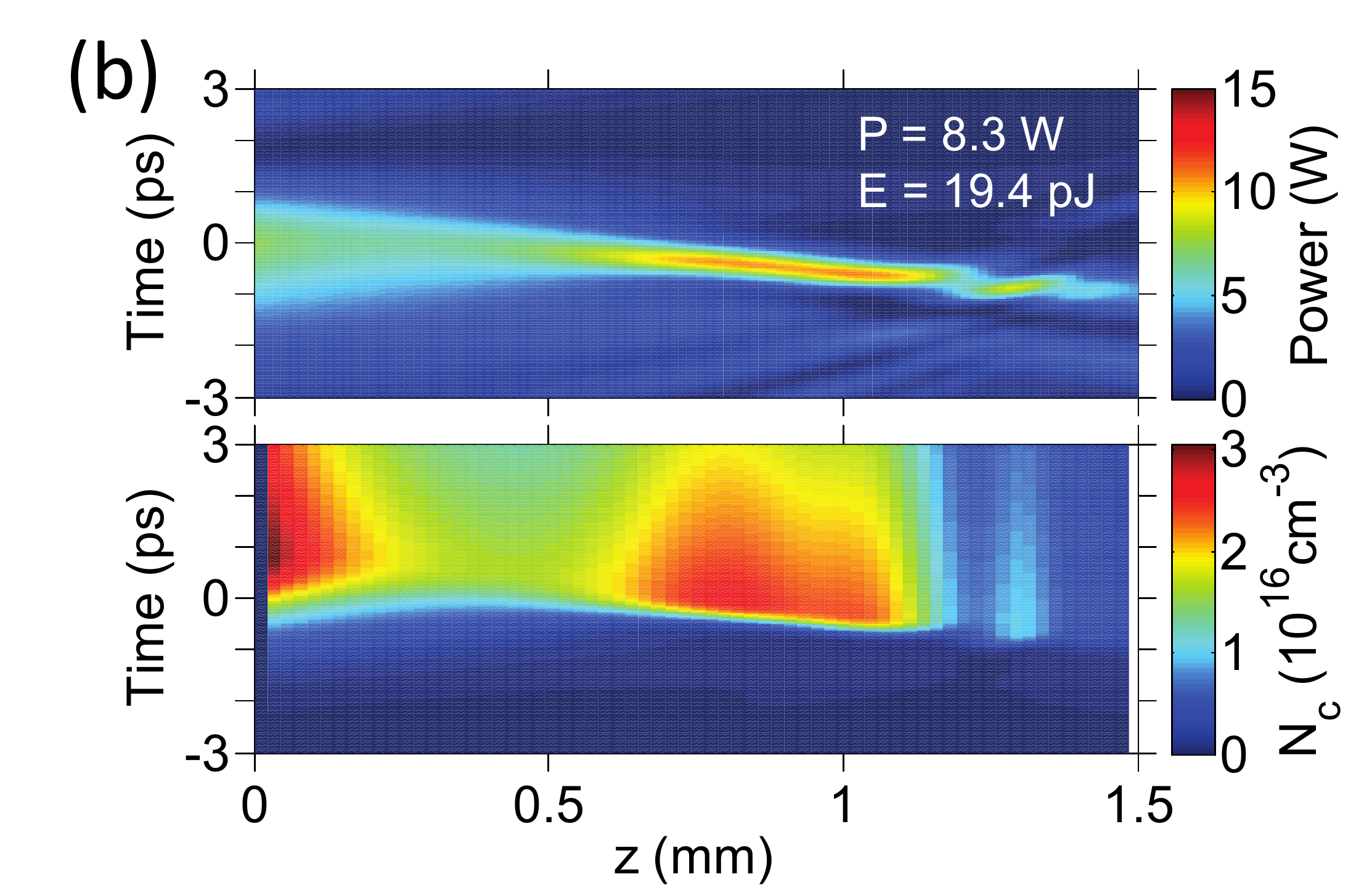} }
\parbox{7.5cm}{\includegraphics[width=7.5cm]{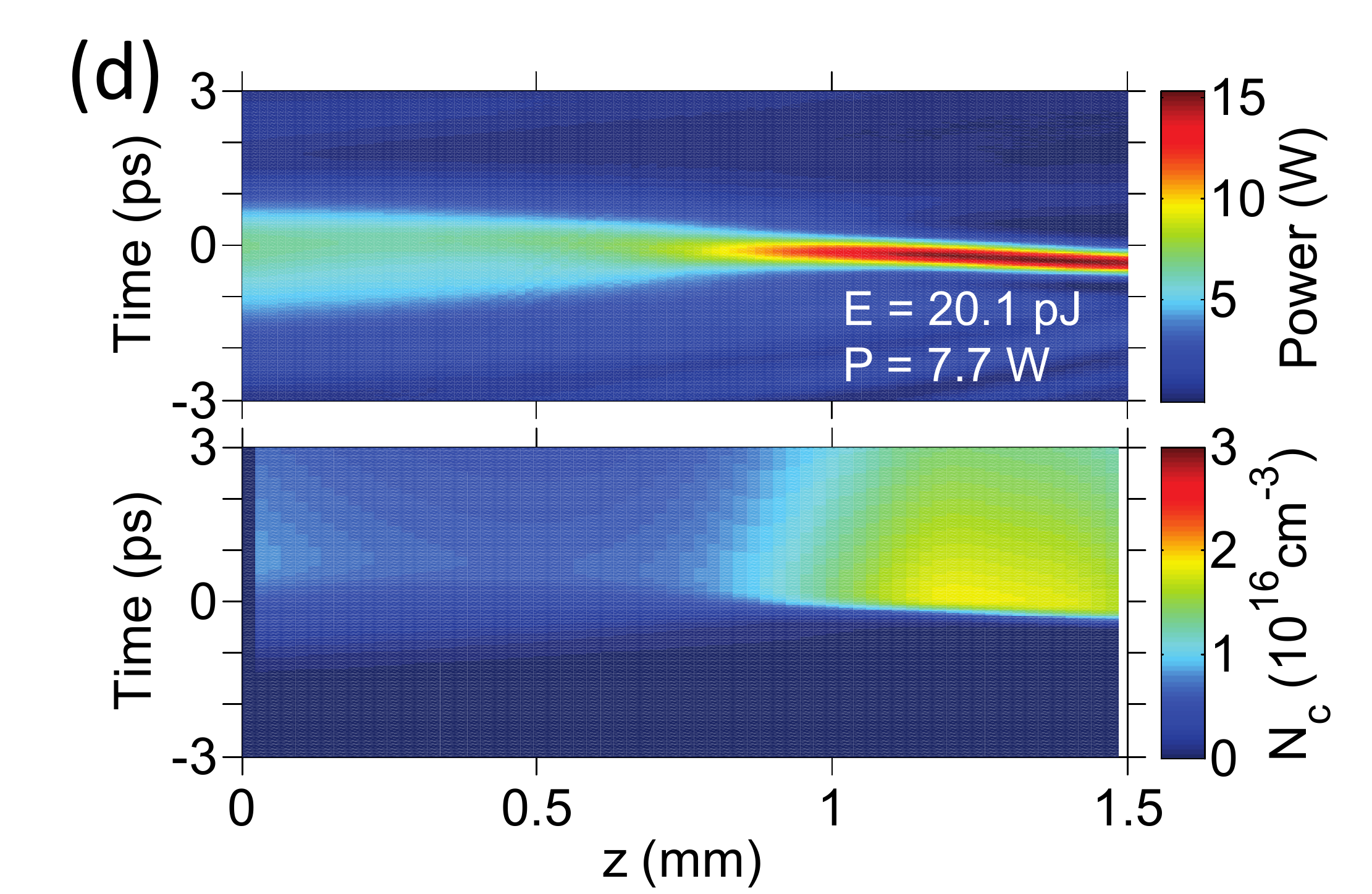} }\\
\caption{Soliton pulse acceleration via self-induced non-adiabatic plasma chirp. (a) NLSE modeled output corresponding to the FROG traces of 1546 nm in Fig. \ref{fig:frogTraces}. The pulse temporally shifts to shorter delays with increased input power, as indicated by the temporal first-order moment (center of mass) of the pulses. The black dashed trace is a numerical simulation with suppressed free-carrier effects ($N_c$=0), demonstrating the shift originates from the generation of a free-carrier plasma. The dashed red line acts a guide to the eyes to the pulse center. Recall that FROG is relative time, e.g. $\tau$ = 0. (b) Pulse intensity and carrier generation along the waveguide length for 1546 nm from NLSE modeling. The role of free-carriers inducing the temporal shift is clearly visible. (c) NLSE and FROG for 1533.5 nm. We measure a minimum temporal duration of 440 fs. (d) The NLSE model indicates less temporal acceleration at 1533.5 nm due to weaker free-carrier effects compared to the slower light at 1546 nm.}
\label{fig:pulseTimeShift}%
\end{figure}
\clearpage
 
\begin{figure}
\centering
\includegraphics[width=12cm]{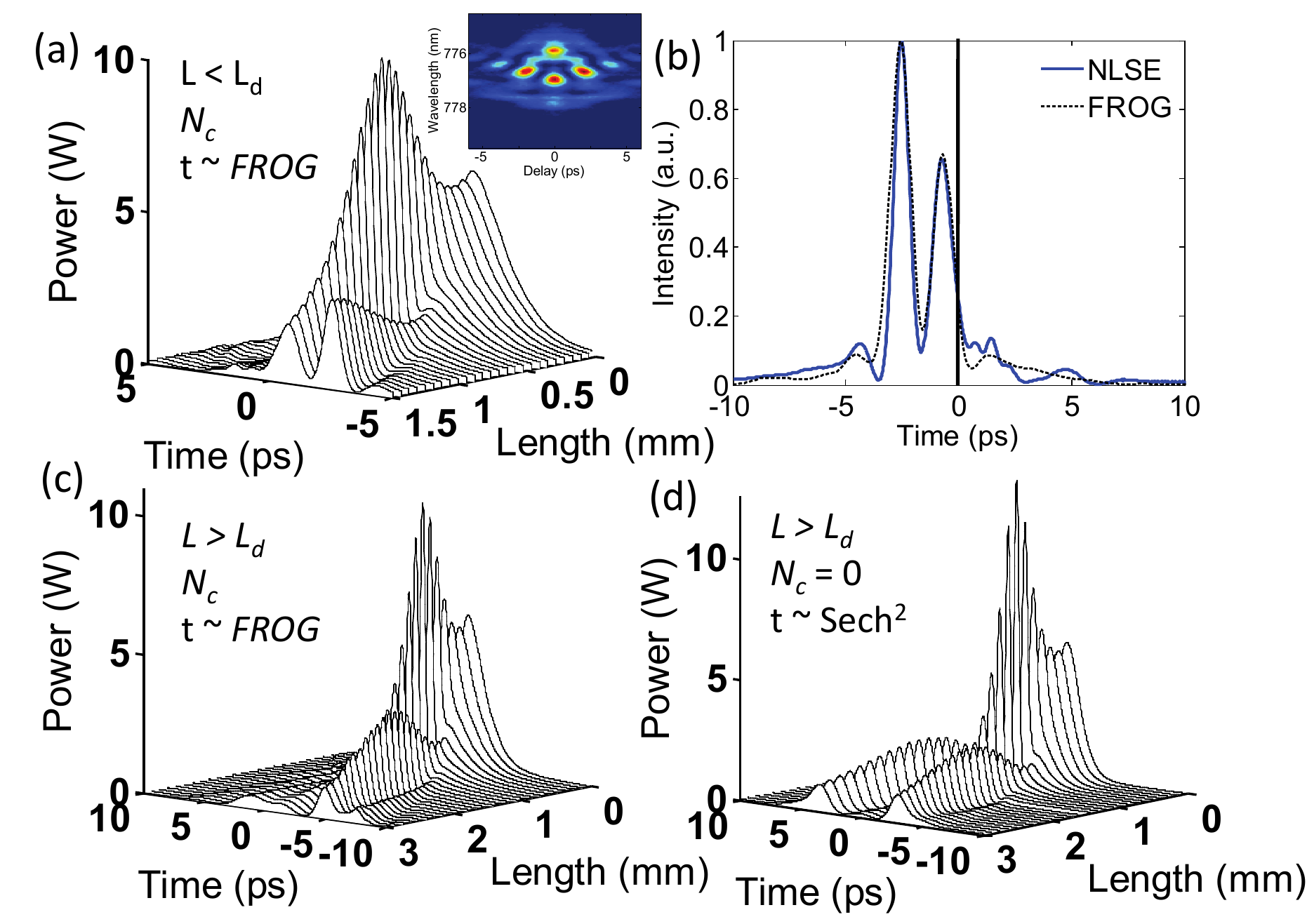}
\caption{Suppression of soliton periodic recurrence in a free-carrier plasma. (a) Waterfall plot of soliton evolution from NLSE at 1555.0 nm (6.3 W, 14.8 pJ) including auxiliary free-carrier non-instantaneous response. Inset: corresponding FROG trace. (b) Comparison of the experimental pulse shape at 1555 nm with the NLSE model. (c) NLSE simulations with parameters identical to (b), except for twice the sample length. (d) NLSE simulations without free-carriers ($N_c$=0) show hints of periodic evolution. The contrast of (d) with panel (c) demonstrates the suppression of periodic recurrence in the presence of free-carrier dynamics.}
\label{fig:carrierEffects} 
\end{figure}
 
\clearpage

\begin{figure}[*h]
\centering
\parbox{7.5cm}{\includegraphics[width=7.5cm]{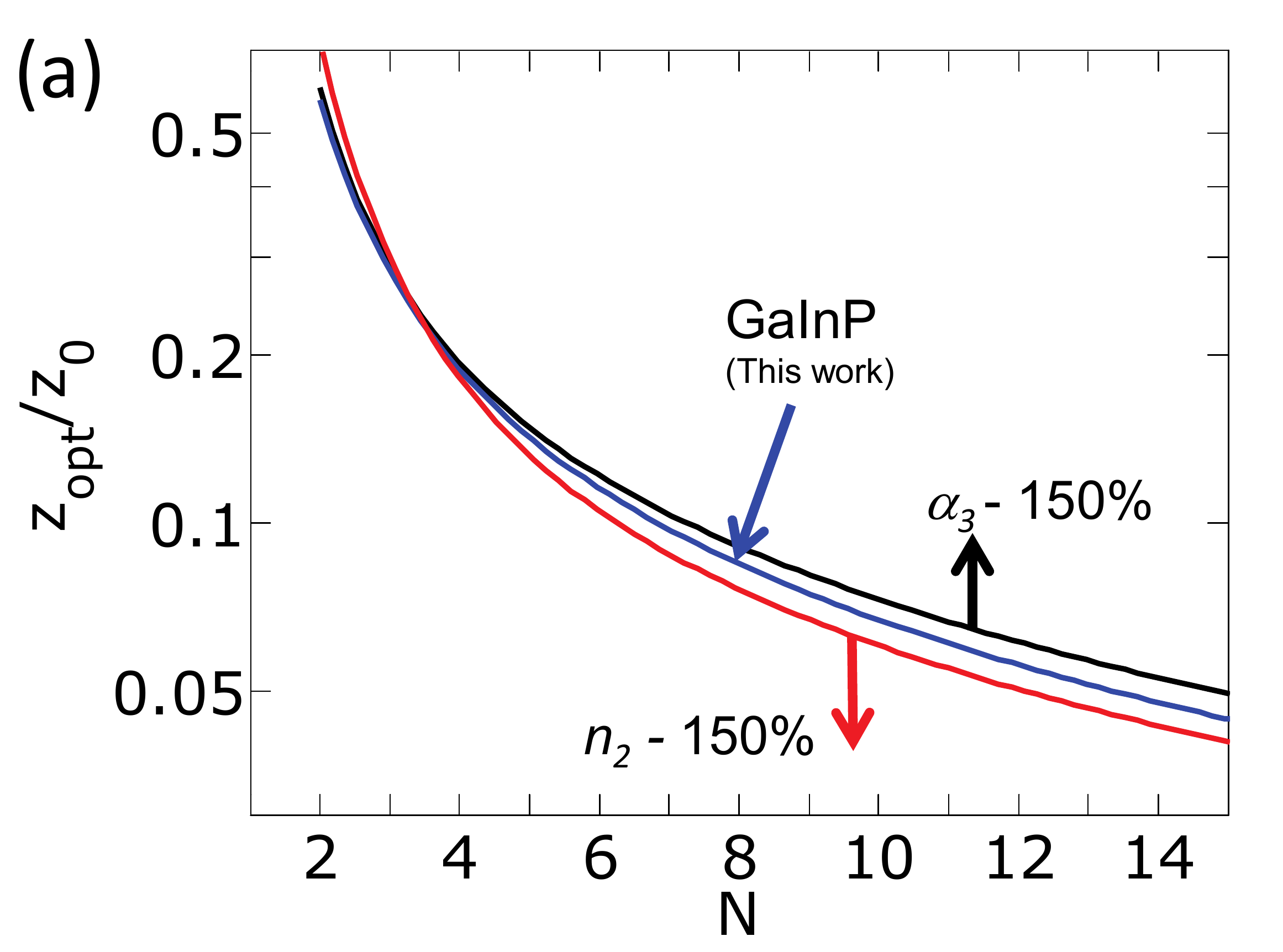}   }
\parbox{7.5cm}{\includegraphics[width=7.5cm]{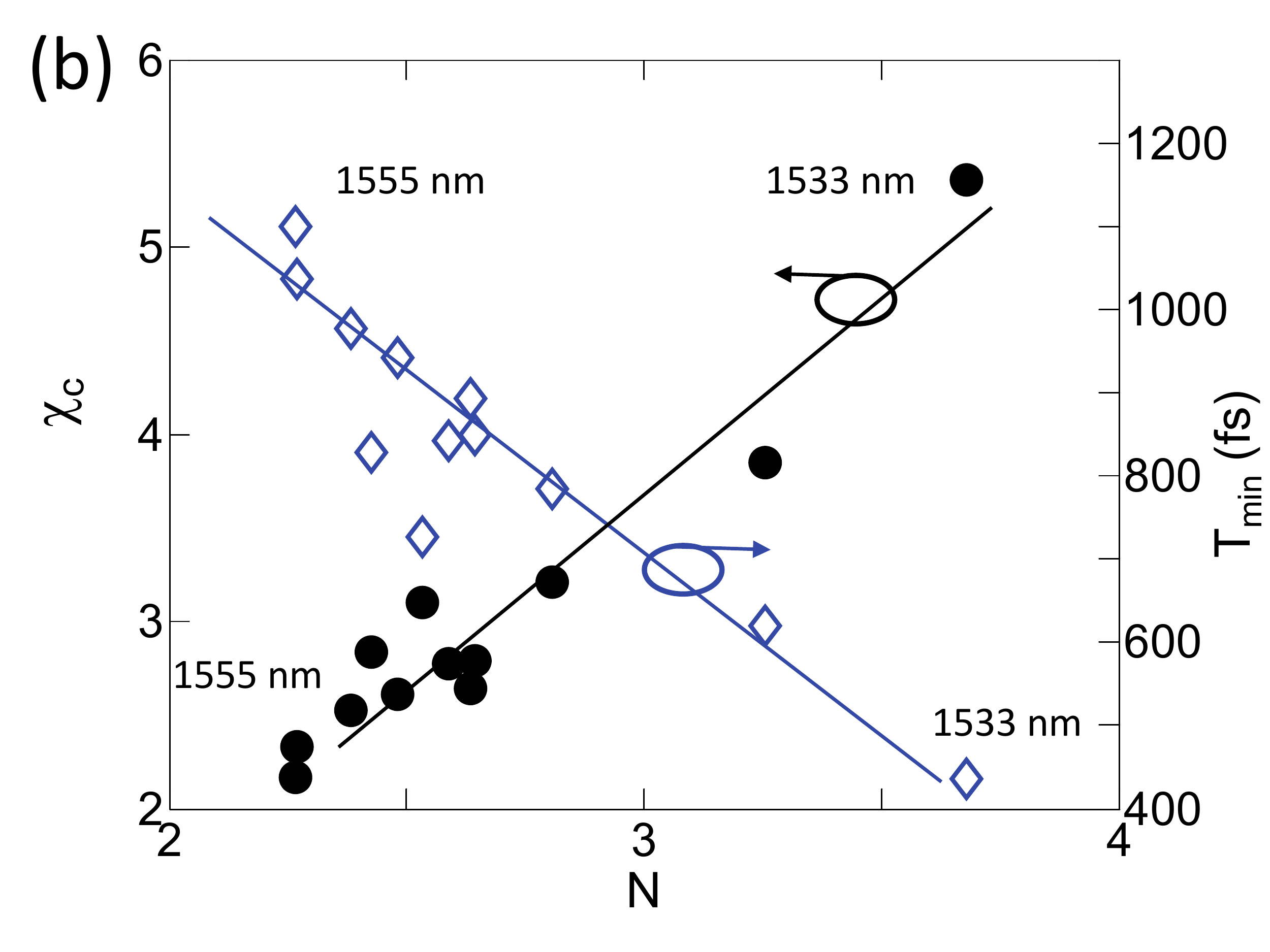} }
\caption{Optimal soliton temporal compression. (a) The ratio $z_{opt}/z_0$ for soliton compression in multiphoton absorption and plasma materials, here specified for three-photon absorption (solid blue). We derive the GaInP dispersive photonic crystal curve from the NLSE. The role of larger effective nonlinearities $n_{2eff}$ and $\alpha_{3eff}$ are indicated by arrows.
(b) Left: Experimentally observed compression factor, $\chi_c$, versus soliton number $N$ obtained at the minimum pulse duration for each wavelength. Right: minimum pulse durations at the various wavelengths ($N$ values) approach 440 fs from the 2.3 ps pulse input at 1533.5 nm.}
\label{fig:minimumDuration}%
\end{figure}

\clearpage
\section*{Supplementary Information}
\setcounter{subsection}{0}                      
\setcounter{figure}{0}                          
\renewcommand{\thefigure}{S\arabic{figure}}     

\subsection*{Linear properties of the photonic crystal waveguide}
\begin{figure}[*h]%
\centering
\includegraphics[width=8cm]{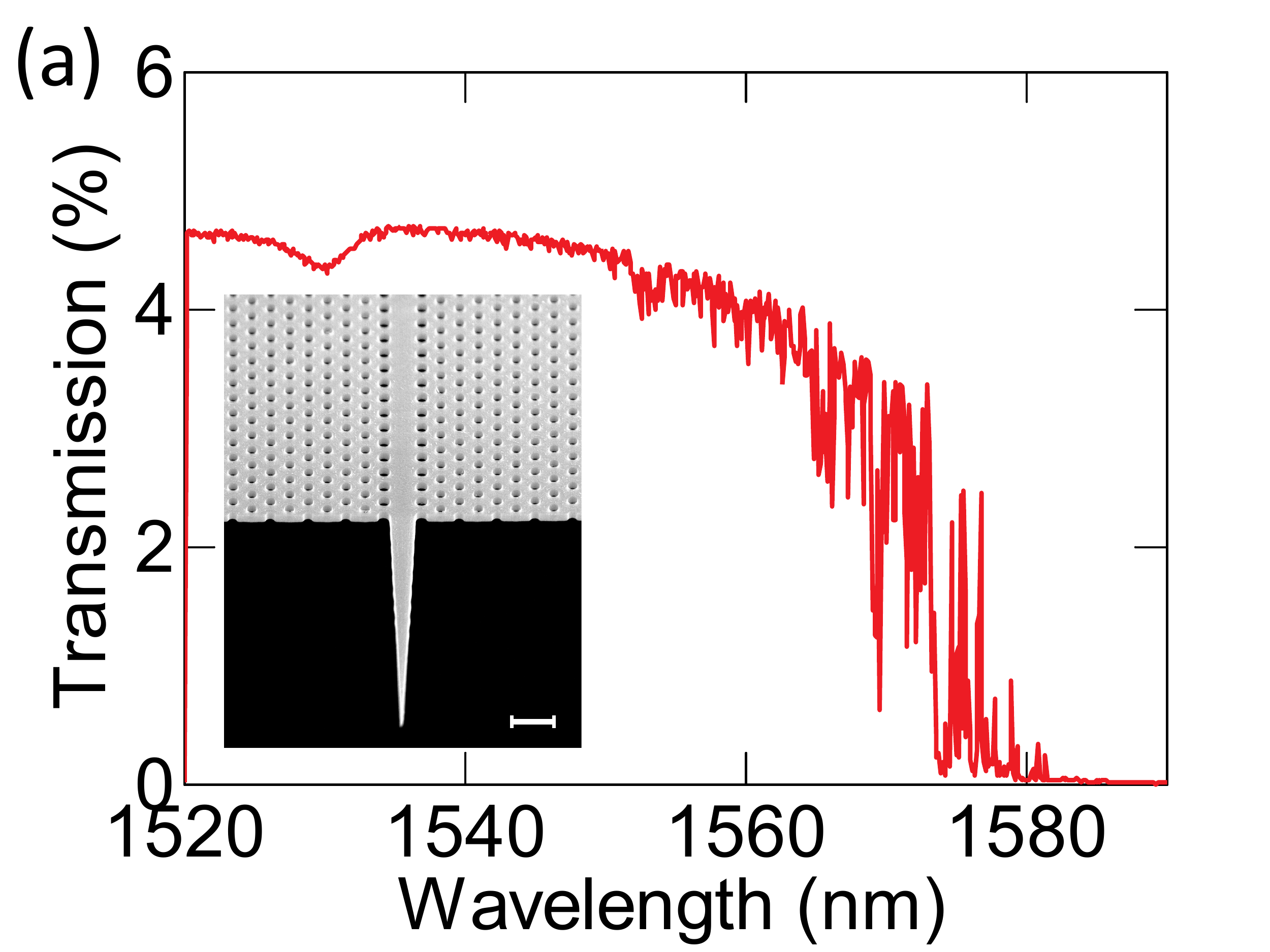}
\includegraphics[width=12cm]{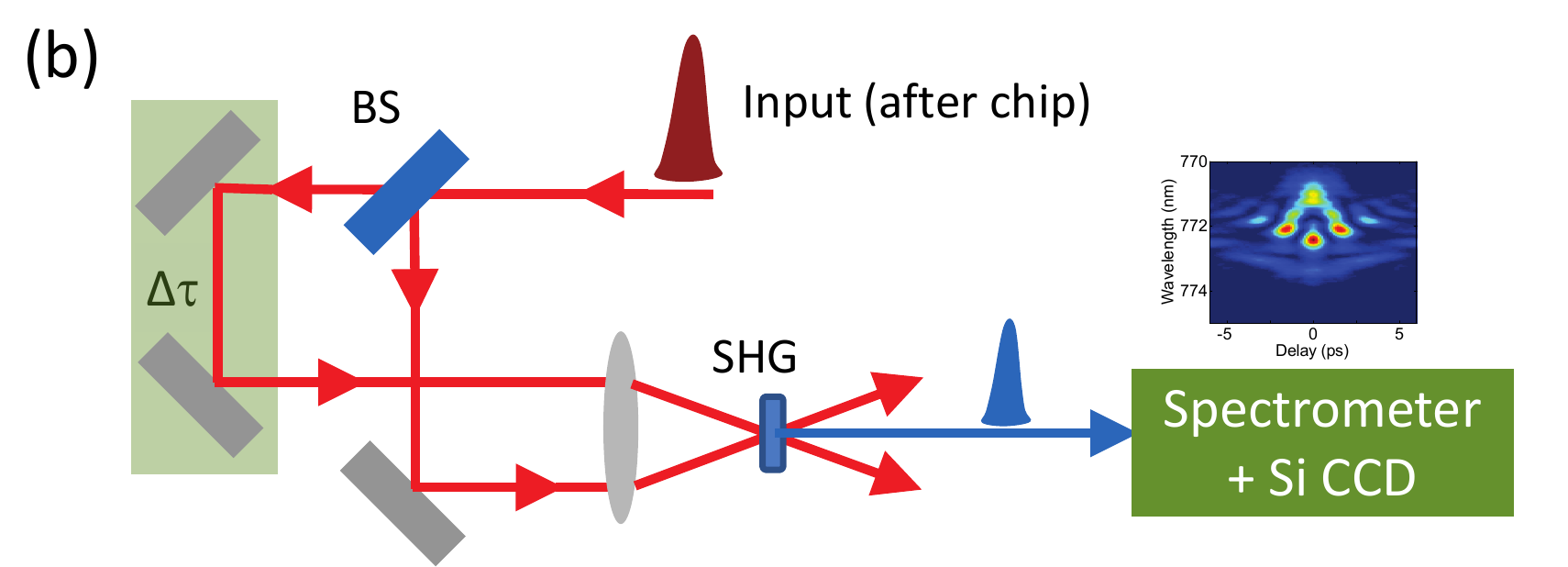}
\caption{Linear properties and home-built FROG setup. (a) Linear transmission of the photonic crystal waveguide device. The dip around 1530 nm is the onset of a higher-order mode, outside the regime of interest. (b) Frequency-resolved optical gating (FROG) setup used to characterize the soliton pulse dynamics, including complete intensity, duration, and phase information. BS: Beam splitter, SHG: BBO second-harmonic crystal, $\Delta\tau$: delay stage.}
\label{fig:linearPropertiesFROG}
\end{figure}

The transmission of the 1.5-mm photonic crystal (PhC) waveguide is illustrated in Fig. \ref{fig:linearPropertiesFROG}(a). Total insertion loss (before and after coupling optics) is estimated to be 13 dB at 1530 nm (group index $n_g$ = 5), including 10 dB attributable to the coupling optics, and 1 dB propagation loss at this wavelength. Carefully designed integrated mode-adapters reduce waveguide coupling losses to 2 dB (insertion) and suppress Fabry-Perot oscillations from facet reflections as shown in the inset of Fig.~\ref{fig:linearPropertiesFROG}(a) \cite{tranAPL09}. The linear loss is $\alpha$ = 10 dB/cm at 1540 nm, scaled linearly with $n_g$ \cite{baba_nature2008,monat2009}. The small feature at 1530 nm is the onset of the higher-order waveguide mode coupling. The energy coupled into the PhC is estimated by assuming symmetric coupling loss (input and output) except for a factor accounting of mode mismatch on the input side (lens to waveguide) that we do not have at the output since $P_{out}$ is measured with a free space power meter. This enables us to calculate the factor between the measured average power at input (output) and the value of the average power at the beginning (end) of the waveguide. Pulse energy is obtained by dividing by the repetition ratio. As noted in the main body, a slight dip is present in the group index at $\sim$ 1545 nm, implying a small deviation in the local dispersion $\beta_2$. This gives rise the the spreading near $N$=2.5 in Fig. \ref{fig:minimumDuration}(b) corresponding to that wavelength region.
 
\subsection*{Frequency-resolved optical gating (FROG) pulse}
\begin{figure}
\centering
\includegraphics[width=12cm]{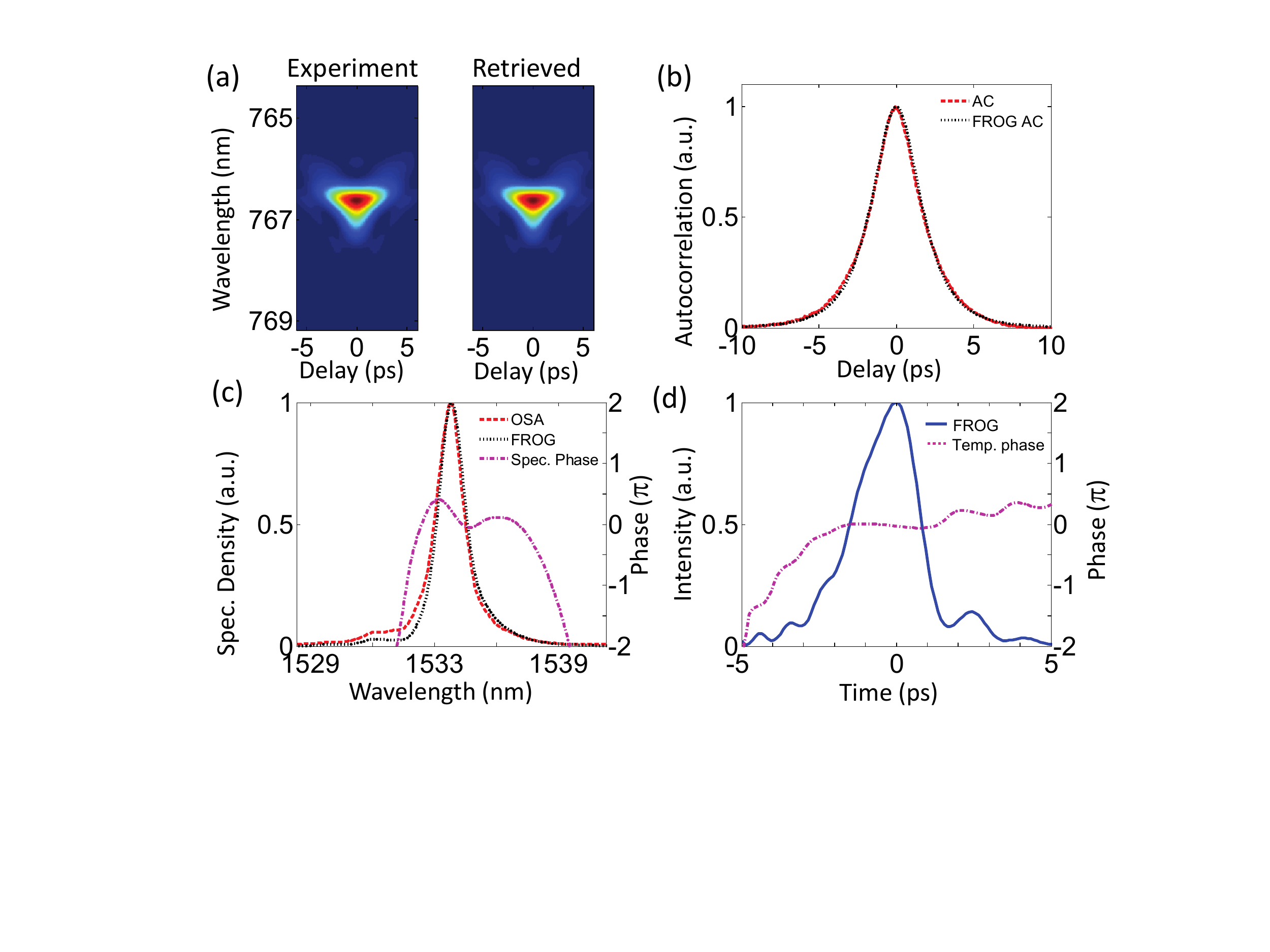}
\caption{Typical input pulse measured by the FROG. Though 1533.5 nm is shown here, other wavelengths exhibit similar characteristics. (a) Experimental and retrieved FROG traces (b) Autocorrelation - FROG (black dotted) and autocorrelator (red dashed) (c) Spectral density FROG (black dotted), optical spectrum analyzer (red dashed) and spectral phase (dash-dot magenta) (d) Temporal intensity (solid blue) and phase (dash-dot magenta).}
\label{fig:figure_frogInput} 
\end{figure}

Fig. \ref{fig:linearPropertiesFROG}(b) shows the frequency-resolved optical gating (FROG) setup used in the experiments. With the FROG technique, one is able to completely characterize the pulse, including intensity and phase information in both the spectral and temporal domains. We employed a second-harmonic FROG (SHG FROG) technique detailed in the Methods. The equation governing the second-harmonic generation SHG-FROG is: 
\begin{equation}
I_{FROG}(\omega, \tau) = \left|\int_{-\infty}^\infty E(t)E(t-\tau)e^{-i\omega t} dt\right|^2,
\end{equation}
where $I_{FROG}(\omega, \tau)$ is the measured pulse, $E(t)$ is the electric field and $e^{-i\omega t}$ the phase. The spectrograms are processed numerically to retrieve the pulse information \cite{trebinoBook}. Fig. \ref{fig:figure_frogInput}(a) compares the experimental and retrieved spectrograms of typical input pulse measured by the FROG, here at 1533.5 nm. Figs. \ref{fig:figure_frogInput}(b) and (c) indicate the FROG autocorrelation and spectrum compared with independent measurements with a commercial autocorrelator (Femtochrome) and optical spectrum analyzer (OSA), respectively. Fig. \ref{fig:figure_frogInput}(d) shows the temporal intensity and phase retrieved from the FROG measurement, information unavailable from typical autocorrelation and OSA measurements. The pulse phase is flat across the pulse, indicating near-transform limited input pulses.
 
\subsection*{Frequency-resolved optical gating of chip-scale ultrafast solitons at 1533.5 nm and 1546 nm}
\begin{figure}[*h]%
\centering
\includegraphics[width=16cm]{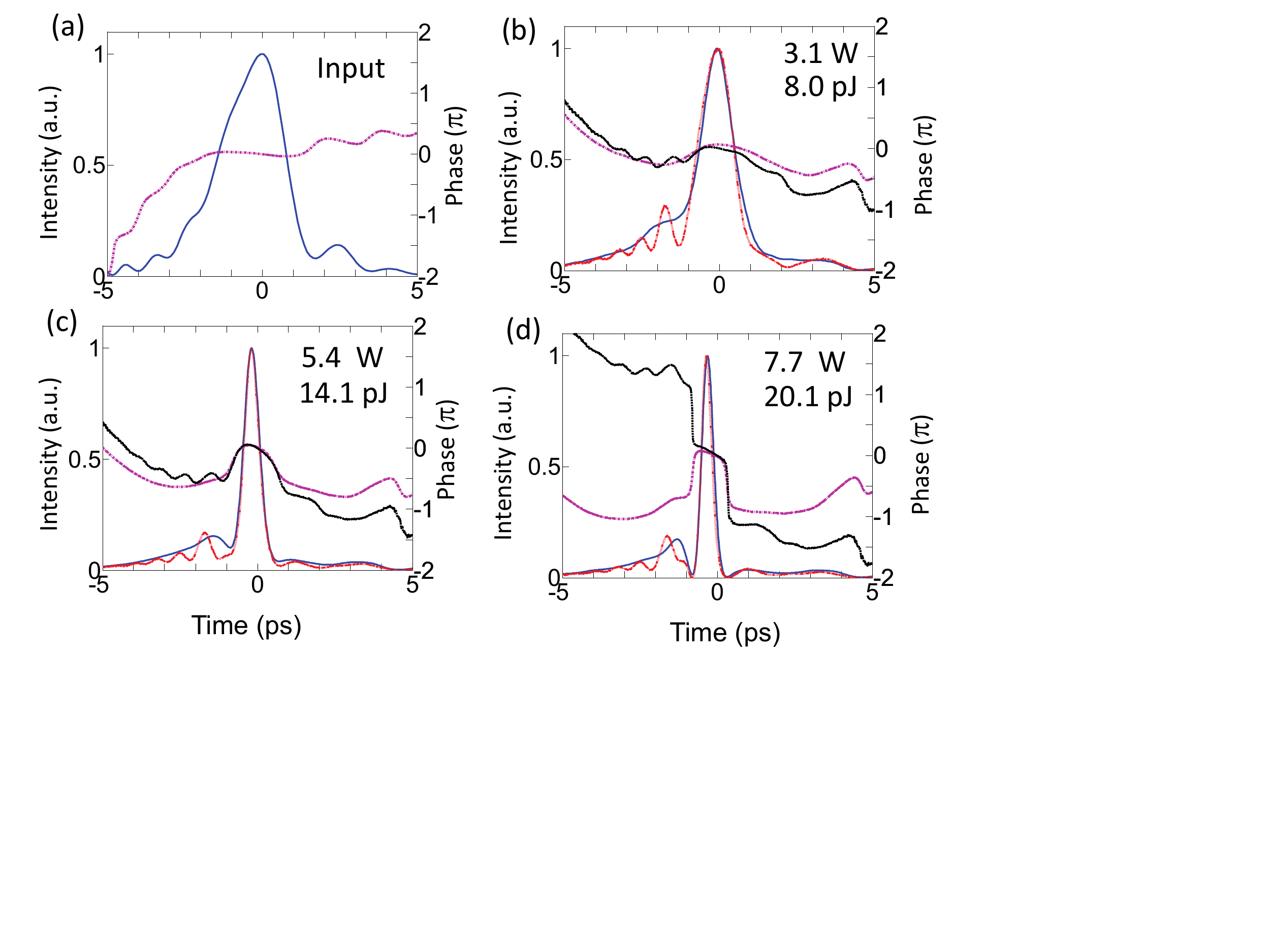}
\caption{Ultrafast soliton compression at 1533.5 nm. Panels (a)-(d) correspond to the spectrograms in Figs. \ref{fig:frogTraces}(i)-(l) in the main text. (a)-(d): FROG retrieved time domain intensity (solid blue) and phase (dashed magenta), with gating error less than 0.005 on all runs. Superimposed nonlinear Schr\"odinger equation modeling: intensity (dashed red), and phase (dash-dot black), demonstrates strong agreement with experiments. Panel (d): The pulse compresses from 2.3 ps to a minimum duration of 440 fs ($\chi_c$ = 5.3) at 20.1 pJ (7.7 W), demonstrating higher-order soliton compression.}
\label{fig:frogTraces_1533nm}
\end{figure}

Fig. \ref{fig:frogTraces_1533nm} shows the retrieved FROG intensity (blue line) and phase (magenta) at 1533.5 nm ($n_g$ = 5.4, $\beta_2$=-0.49 ps$^2$/mm). The nonlinear Schr\"odinger equation results are presented in Figs. \ref{fig:frogTraces_1533nm} (a)-(d) with predicted intensity (dashed red) and phase (dash-dot black). Since FROG only gives the relative time, we temporally offset the FROG traces to overlap the NLSE for direct comparison. All parameters precisely determined from experimental measurements, e.g. no free fitting parameters. Fig. \ref{fig:frogTraces_1533nm}(d) shows the maximum pulse compression to a minimum duration of 440 fs from 2.3 ps ($\chi_c$ = 5.3) at 7.7 W (20.1 pJ, $N$ = 3.5), demonstrating higher-order soliton compression. The slight dip in the pulse phase at positive delay (temporal tail) is due to free-carrier blue-shift. 

\begin{figure}[*h]%
\centering
\includegraphics[width=14cm]{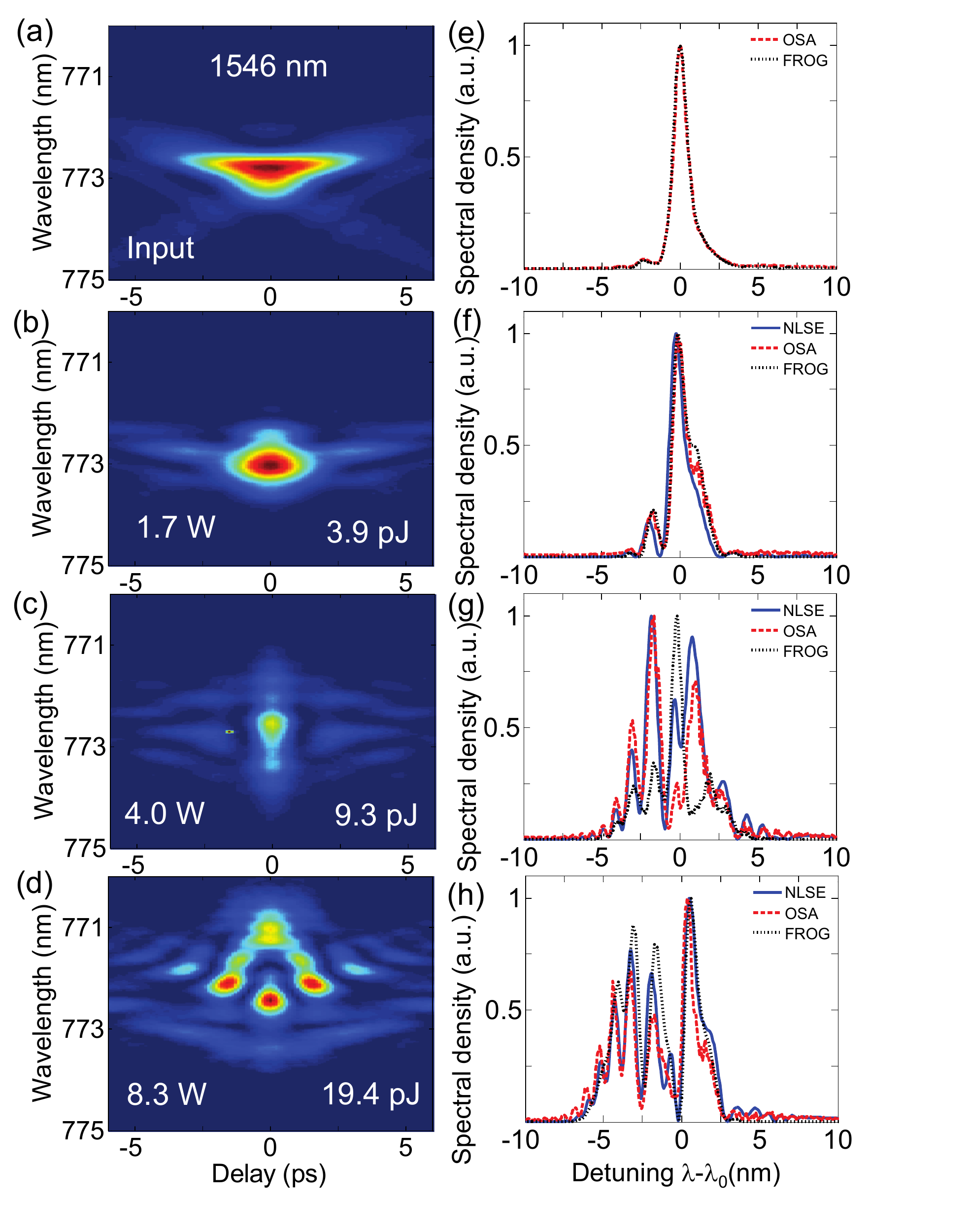}
\caption{Spectral properties of pulses at 1546 nm. (a)-(d): FROG spectrograms with coupled pulse energies from 3.9 pJ to 20.1 pJ repeated from the main text for simple comparison. (e)-(h): FROG retrieved spectral density (dashed black ), OSA (dashed red), and superimposed NLSE modeling (solid blue) demonstrate agreement in both the spectral domain (shown here) and time domain (main text).}
\label{fig:frogOSA_1546nm}
\end{figure}

Figs. \ref{fig:frogOSA_1546nm}(a)-(d) show the FROG traces at 1546 nm as in the main paper. Figs. \ref{fig:frogOSA_1546nm}(e)-(h) compare the retrieved FROG spectral density (dotted black) and NLSE simulations (solid blue) to independent measurements with an optical spectrum analyzer (dashed red). The experimental and modeling results agree simultaneously in both the time (main text) and spectral domains shown here.

\subsection*{Periodic soliton recurrence and suppression in the presence of free-electron plasma: role of free-carriers and input pulse shape}
\begin{figure}[*h]%
\centering
\includegraphics[width=10cm]{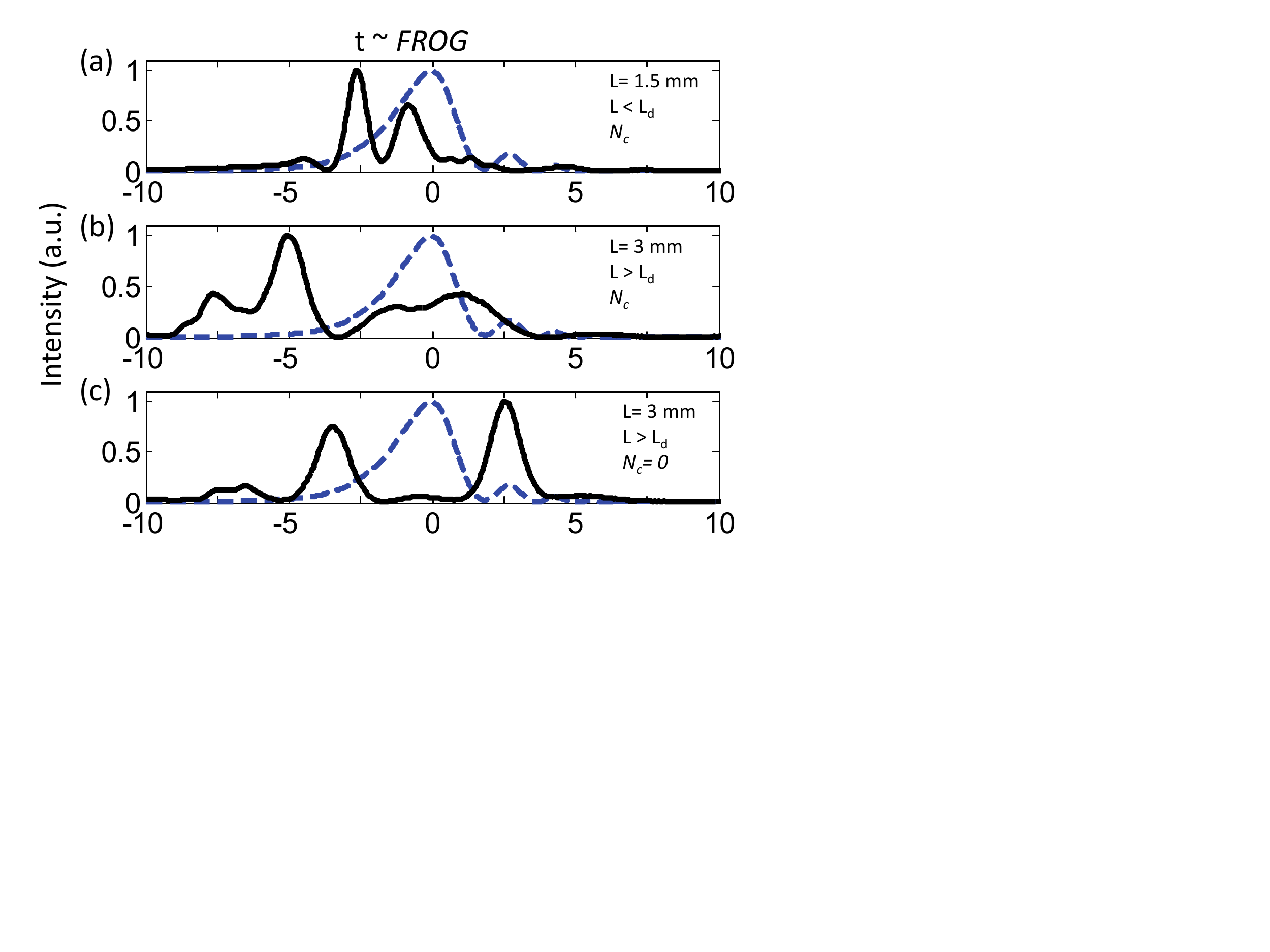}
\vspace{0.5cm}
\includegraphics[width=10cm]{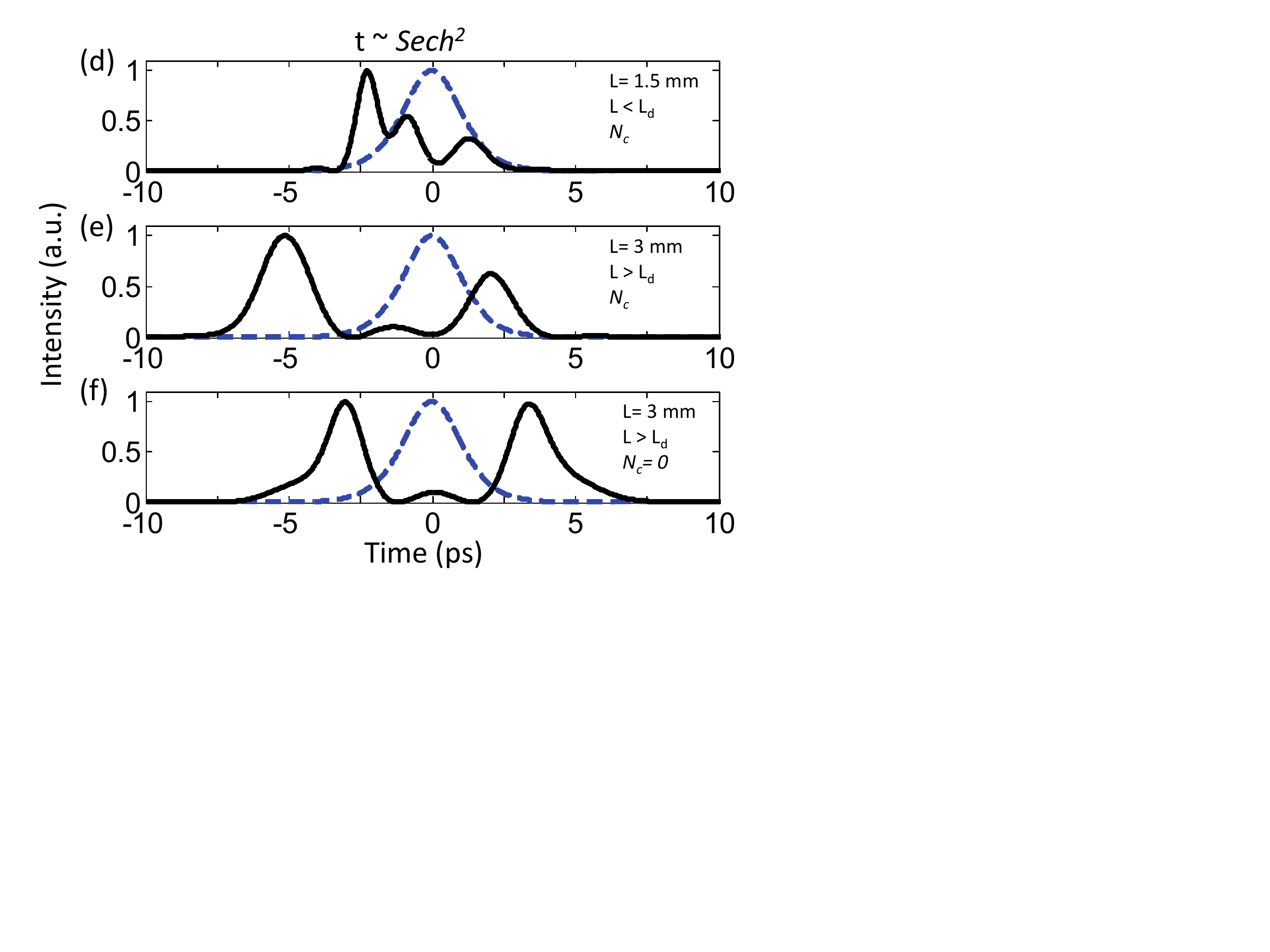}
\caption{Suppression of soliton periodic recurrence: role of free-carriers and input pulse shape. Panels (a)-(c): NLSE with experimental FROG input pulse. (a) Full simulation $L$ = 1.5 mm and free carriers ($N_c$) as in the main paper. (b) $L$ = 3 mm with free-carriers ($N_c$). (c) $L$ = 3 mm with suppressed free-carriers ($N_c=0$). Panels (d)-(f), same as (a)-(c) with NLSE with $sech^2$ input pulse.}
\label{fig:comparisonTimeDomainNLSE}
\end{figure}

In Fig. \ref{fig:carrierEffects} of the main text, we demonstrated the suppression of periodic soliton recurrence in the presence of free-electron plasma. Fig. \ref{fig:comparisonTimeDomainNLSE} shows additional details of the physics presented there. Fig. \ref{fig:comparisonTimeDomainNLSE}(a) shows the NLSE model of the experimental situation: $L$ = 1.5 mm and free carriers ($N_c$) as in the main paper. Fig. \ref{fig:comparisonTimeDomainNLSE}(b) shows that even with longer $L$ = 3 mm samples the pulse recurrence is clearly suppressed. Fig. \ref{fig:comparisonTimeDomainNLSE}(c) shows NLSE modeling in the absence of free-carriers ($N_c=0$). The pulse splits temporally, but does not reform due to loss. In contrast to the FROG input pulses used in the simulations throughout the text thus far, Figs. \ref{fig:comparisonTimeDomainNLSE}(d)-(f) show NLSE models with chirp-free $sech^2$ input pulses. Importantly, the same basic features are represented for both the FROG (a)-(c) and $sech^2$ inputs (d)-(f), demonstrating soliton re-shaping of our experimental pulses.

\subsection*{Pulse acceleration in a multiphoton plasma}
\begin{figure}[*h]%
\centering
\includegraphics[width=10cm]{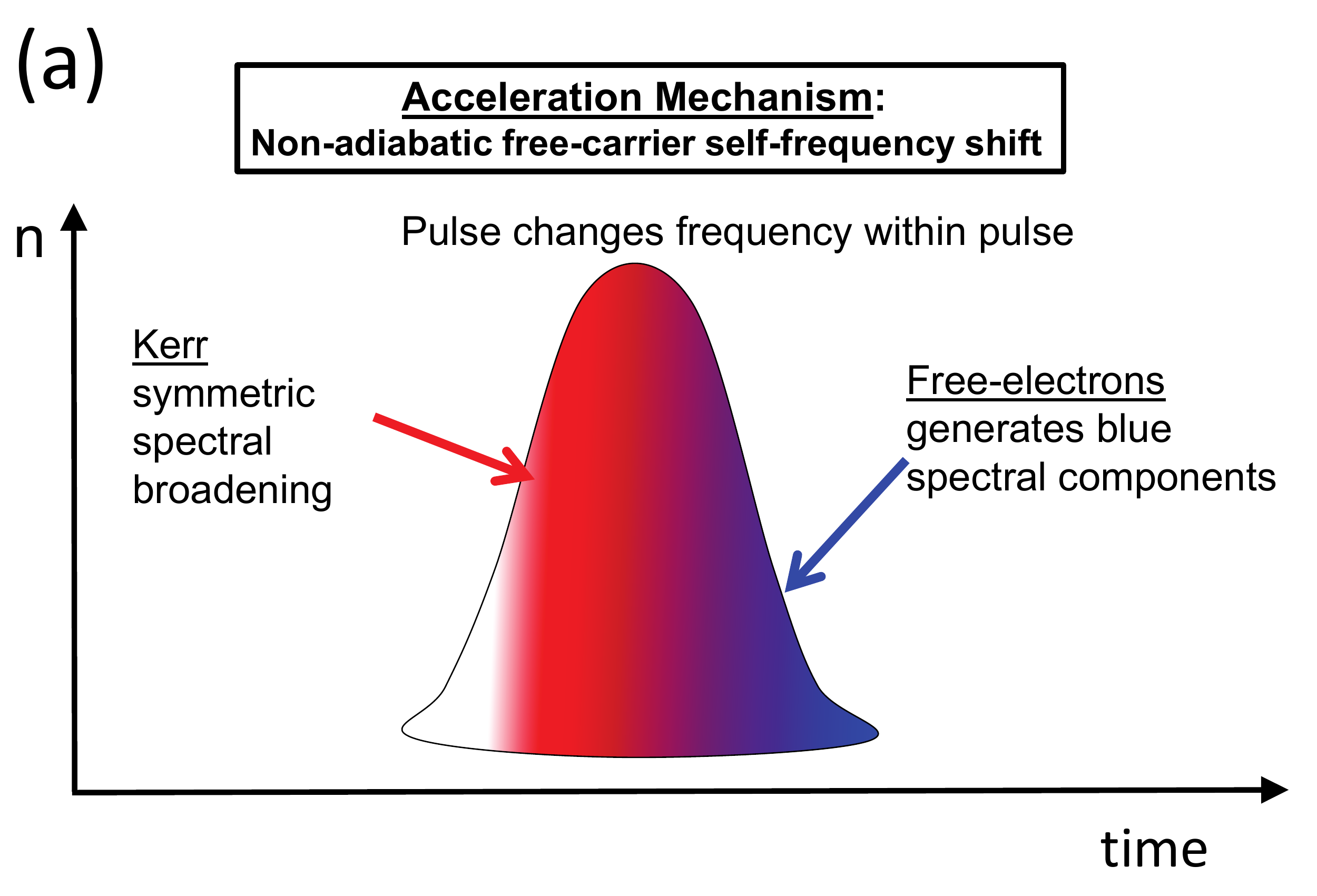}
\includegraphics[width=10cm]{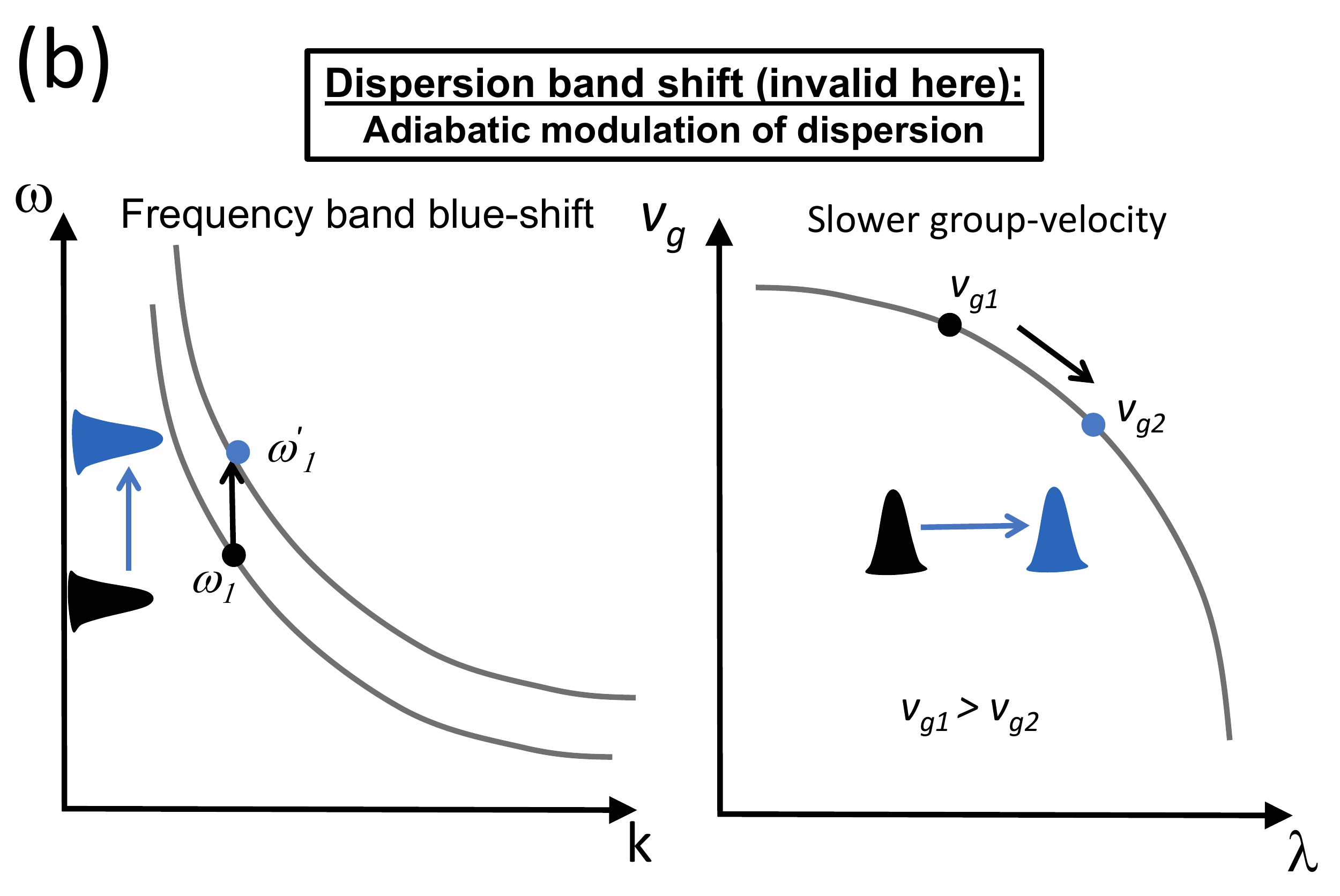}
\caption{Pulse modulation mechansisms. (a) Acceleration is due to the self-induced frequency-chirp due to non-adiabatic free-carrier generation within the pulse. This is confirmed via the NLSE simulations in the paper. (b) Dispersion band-shift due to adiabatic modulation of free-carriers at large intensities induces a frequency conversion process. This is not the case here.}
\label{fig:bandShift}
\end{figure}

The mechanism accelerating the pulse is a non-adiabatic generation of a free-carrier plasma via multiphoton absorption within the pulse inducing a blue frequency chirp. Fig. \ref{fig:bandShift}(a) shows a schematic of the self-induced free-carrier blue-shift and resulting acceleration of the pulse. The regions of largest plasma generation occur at the waveguide input as well as at points of maximum compression as shown in Fig. \ref{fig:pulseTimeShift} of the main text. Moreover, we note that the dispersion bands themselves do not shift at our 10-pJ 1550-nm pulse energies, in contrast to other reports with Ti:sapphire pump-probe and carrier injection with above-band-gap 1 to 100 nJ pulse energies at $\sim$ 800-nm~\cite{ kampfrath2010ultrafast,leonard2002ultrafast}. Such a scenario, presented in Fig. \ref{fig:bandShift}(b), would only cause the light to shift slower group velocities, as has been shown in Ref. \cite{kampfrath2010ultrafast}. Furthermore, this mechanism is not a deceleration, but rather a frequency conversion method to change the pulse central wavelength to a frequency with different propagation properties.

\clearpage
\clearpage
\clearpage
\clearpage
\clearpage
\clearpage

\clearpage

\end{document}